\documentclass[aps,twocolumn,amsmath,amssymb,preprintnumbers,floatfix,prx,superscriptaddress,longbibliography]{revtex4-2}%{revtex4}%

\usepackage{braket}
\usepackage{comment}
\usepackage[version=4]{mhchem}
\usepackage[utf8]{inputenc}
\DeclareUnicodeCharacter{03C0}{\ensuremath{\pi}}
\usepackage{newtxtext}
\usepackage[upint]{newtxmath}
\usepackage{microtype}
\usepackage{textcomp}
\usepackage{dsfont}
\usepackage{eucal}
\usepackage{siunitx}
\usepackage{soul}
\usepackage[normalem]{ulem}
\usepackage{enumerate}
\usepackage{amsfonts}
\usepackage{color}
\usepackage{soul}

\usepackage{todonotes}
\presetkeys%
    {todonotes}%
    {inline}{}

\usepackage{graphicx}

\usepackage[colorlinks,allcolors=blue]{hyperref}
\usepackage[capitalize]{cleveref} % Do we want to capitalize?
\usepackage{cleveref}

\renewcommand{\v}[1]{\boldsymbol{#1}}                           % Bold vector
       
\newcommand{\fermi}{\text{F}}
\definecolor{DarkBlue}{rgb}{0,0,0.80}
\definecolor{DarkRed}{rgb}{0.80,0,0}
\definecolor{Purple}{rgb}{0.55,0,0.55}
\definecolor{Purple}{rgb}{0,0,0.8}

\renewcommand{\H}[1]{\hat{#1}}

\newcommand{\B}[1]{\bm{#1}}

\newcommand{\A}[1]{\H{#1}^A}

\newcommand{\eg}{\textit{e.g.}\ }

\begin{document}

% \title{Enhanced qubit performance by integrating altermagnets into superconducting qubit designs}
\title{Superconducting Qubits with Altermagnetic Josephson Junctions}
\author{Johanne Bratland Tjernshaugen}
\affiliation{Center for Quantum Spintronics, Department of Physics, Norwegian \\ University of Science and Technology, NO-7491 Trondheim, Norway}
\author{Morten Amundsen}
\affiliation{Center for Quantum Spintronics, Department of Physics, Norwegian \\ University of Science and Technology, NO-7491 Trondheim, Norway}
\author{Jacob Linder}
\affiliation{Center for Quantum Spintronics, Department of Physics, Norwegian \\ University of Science and Technology, NO-7491 Trondheim, Norway}\date{\today}
\begin{abstract}
Identifying a materials platform for creating qubits that are both tunable and resilient towards environmental noise is one of the main hurdles that need to be overcome to realize quantum computation that is practically useful. One pursued avenue to this end is to use superconducting qubits with intrinsic spin-dependent interactions, such as spin-orbit coupling or magnetism. However, the recently discovered class of materials known as altermagnets remains largely unexplored in this context. We here use microscopic calculations to determine how the properties of superconducting qubits are modified when altermagnetic Josephson junctions are included. The key qubit performance parameters, including splitting, anharmonicity, decoherence, and single/coupled-qubit gate operation times, display rich behavior depending on the characteristic properties of the altermagnetic material, such as the strength of the Néel field and the crystallographic orientation of the altermagnetic relative to the interfaces in the system. We focus in particular on the transmon design and show that the qubit is very well protected against decoherence and simultaneously shows superior anharmonicity both near 0-$\pi$ transition points and when it is in a $\phi$-state. We propose that by using strain, the altermagnetic qubit can be moved out of its protected regime to enable faster gate operation times, and then moved back to its protected state. We establish the physical mechanism underlying the behavior of all central qubit metrics, clarifying how real devices interpolate between an altermagnetic double-well regime exhibiting barrier-induced protection and a conventional transmon-like single-well regime. We also discuss how the altermagnetic properties influence flux qubits and fluxonium. Our results suggest that integration of altermagnetic materials into existing superconducting qubits design can substantially improve their performance due to the unique properties of the altermagnetic band-structure. 
\end{abstract}

\maketitle

\section{Introduction}

Qubits, effective two-level quantum systems, are the building blocks of quantum computers. While there exist many physical systems that can realize qubits, in particular superconducting qubits have garnered much attention when it comes to actual working prototypes of quantum computers \cite{kjaergaard_arcmp_20, jiang_nsr_25, ezratty_epja_23}. The key component in the vast majority of architectures for superconducting qubits is a Josephson junction. This element consists of a non-superconducting interlayer separating two superconductors. It introduces a necessary non-linearity in the system, which provides anharmonicity in the quantum energy levels of the system, and thus allows for projection of a subspace of two energy levels.

Major hurdles nevertheless exist with regard to making useful quantum computers, including miniaturization of qubit designs and, arguably more crucially, ensuring low noise-sensitivity. For instance, whereas flux qubit designs 
boast advantages such as large anharmonicity and magnetic tunability, large-scale implementation of such qubits with non-magnetic materials presents challenges. One such challenge is the necessity of providing an external magnetic field corresponding to half a flux quantum
to achieve an optimal point of operating the device. This can be solved by including a magnetic element \cite{yamashita_prl_05}, which acts as an intrinsic superconducting phase-shifter. This obviates the need for an external flux to reach the point of optimal operation, which was recently experimentally demonstrated \cite{kim_commat_24}. Incorporating magnetic elements in Josephson junctions nevertheless presents challenges of their own. A magnet surrounds itself with a stray field, which will act parasitically on neighboring qubits on a chip. Allowing the magnet to enter a multi-domain state reduces this problem, but in turn hampers the intrinsic phase-shift provided to the Josephson junction.

To this end, a new class of magnetic materials could provide a solution. It has recently been discovered that collinear antiferromagnets can host nonrelativistic spin-split electron
bands in momentum space while simultaneously having zero net magnetization \cite{smejkalprx22a, smejkalprx22b}. This material class has been named 
\textit{altermagnets}, and can be formally defined by means of symmetry operations acting distinctly on spin and
real space in crystals. One can think of altermagnets combining the best of both worlds from ferromagnets and antiferromagnets: void of net
magnetization and the resulting stray field, altermagnets
have spin-polarized electron bands not limited by relativistic physics, unlike spin-orbit coupled systems. For this reason, the spintronics community has embraced altermagnetism as holding potential for new and improved functionality in spin-based quantum devices.
 An altermagnetic phase has been predicted
and experimentally observed in several materials \cite{song_natrevmat_25}.

Using altermagnets in qubits nevertheless remains largely unexplored up to now \cite{vosoughinia_arxiv_25}. Since interesting physical properties that are absent in conventional non-magnetic Josephson junctions, such as $0-\pi$ oscillations \cite{ryazanov_prl_00}, $\phi$-states \cite{buzdin_prb_03}, and very efficient dissipationless diode effects \cite{nadeem_natrevphys_23}, have been found to exist in altermagnetic junctions \cite{ouassou_prl_23, beenakker_prb_23, lu_prl_24,  chakraborty_prl_25}, it is an intriguing prospect to consider how integration of altermagnets into existing superconducting qubit designs would alter their properties. 

\begin{figure}
    \centering
    \includegraphics[width=0.4\textwidth]{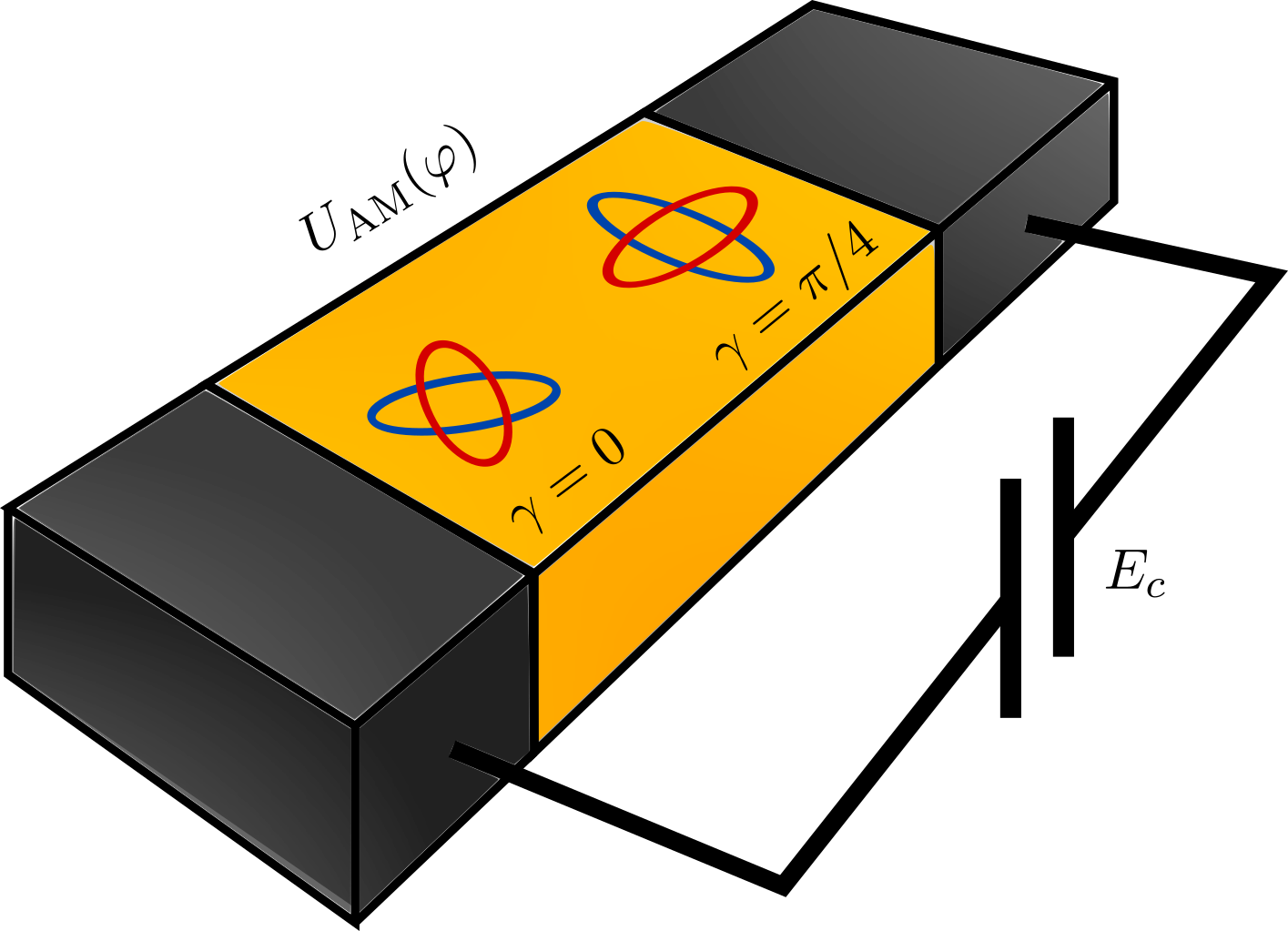}
\caption{We focus on a transmon qubit design where an altermagnetic Josephson junction is parallel-coupled to a capacitor. The JJ has a free energy $U_\text{AM}(\varphi)$ and we consider two crystallographic orientations of the altermagnetic lattice relative the interfaces to the superconductors, depicted in the figure.  
}
    \label{fig:transmon}
\end{figure}

To answer this question, we here perform extensive microscopic calculations to model realistic altermagnetic Josephson junctions. We show that altermagnetic Josephson junctions provide an intrinsic and tunable mechanism for generating double-well superconducting qubits, offering an alternative to superinductors and offering \textit{in-situ} control over the protection-operation tradeoff via their response to strain. Focusing on the transmon design shown in Fig. \ref{fig:transmon}, we show that inclusion of altermagnetic JJs provide an operational regime with both low charge noise sensitivity and large anharmonicity simultaneously. We provide a detailed overview of how key qubit performance parameters such as splitting, anharmonicity, decoherence and dephasing rates, and gate operation times, depend on the properties of the altermagnetic layers. In this way, we identify precisely under which conditions the altermagnetic state yields an enhanced qubit performance. In particular, we propose that altermagnet-based superconducting qubits are suitable for operation near either $0-\pi$ transitions or $\phi$-states, with \textit{in situ} modification via strain engineering. Our results provide guidance for experimental conditions required to design resilient, isolated two-level quantum systems using altermagnets.   

This paper is organized as follows. We start in Sec. \ref{sec:AMJJ} by characterizing the energy-phase relation and its corresponding decomposition into harmonic functions for altermagnetic Josephson junctions, which will turn out to be essential for the qubit performance. We include both an analytical Green function calculation and exact numerical diagonalization results to this end. In Sec. \ref{sec:singlequbit}, we proceed to incorporate altermagnets in a transmon. We compute the characteristic qubit properties as a function of the altermagnetic strength and orientation relative to the interfaces in the Josephson junction, to focus on the unique impact of the altermagnetism. Afterwards, we characterize qubit operation by determining charge and dielectric noise-induced relaxation and dephasing times as a function of the same parameters, determining gate times and how initialization and readout can be performed. While our main focus is a transmon architecture, we briefly discuss the impact an altermagnetic JJ has on a flux qubit and fluxonium design. We then couple two altermagnetic qubits capacitively in Sec. \ref{sec:doublequbit} and characterize the operation of an entangling CZ-gate, which in combination with arbitrary single-qubit motion on the Bloch sphere (discussed in Sec. \ref{sec:singlequbit}) in principle allows for universal quantum computation. We give concluding remarks in Sec. \ref{sec:conclusion} and mention possible future directions. Finally, we include a set of comprehensive appendices which provide a detailed exposition of the theoretical and numerical techniques used to obtain our results.

\section{Altermagnetic Josephson junction}\label{sec:AMJJ}
Altermagnetic Josephson junctions (JJs) have been intensively studied theoretically in recent years, unveiling key features such as $0-\pi$ oscillations \cite{ouassou_prl_23, beenakker_prb_23, zhang_natcom_24} and the emergence of a $\phi$-state \cite{lu_prl_24, fukaya2025josephson}. Both of these properties are relevant for functionality in various qubit designs. The $0-\pi$ transition phenomenon refers to the property that the ground-state phase difference for the JJ can be either 0 or $\pi$, depending on the junction parameters. These parameters include the length of the junction, the altermagnetic strength or the orientation of the altermagnetic band-structure lobes relative to the interfaces. It has been noted \cite{yamashita_prl_05} that the design of flux qubits incorporating one 0 and one $\pi$-junction allows for operation at zero external flux, a proposition that was very recently experimentally realized \cite{kim_commat_24}. The $\phi$-state \cite{buzdin_prb_03}, on the other hand, describes a situation where the Josephson energy has a doubly degenerate minimum, rather than a single minimum at either 0 or $\pi$. We proceed to show that the existence of 0, $\pi$, and $\phi$-states in altermagnets provides the opportunity to tailor the JJ energy potential, which in turn drastically affects all of the qubit functional properties.

\subsection{Analytical solution: $0-$, $\pi-$, and $\phi-$ states} 
When an electron travels through a barrier separating two reservoirs, it accumulates a phase $\psi=kd$, where $k$ is its wave vector and $d$ the barrier length. In a Josephson junction, such a phase is important as it directly influences the current-phase relation, and in turn the energy-phase relation. Furthermore, spin-dependent barriers give rise to a spin-dependent phase accumulation, and this phase difference $\Delta\psi=\psi_\uparrow-\psi_\downarrow$ between electrons of opposite spins has observable consequences. In a ferromagnetic barrier, it modulates the Josephson current by $\cos\Delta\psi$ to lowest order in the tunneling amplitude, with $\Delta\psi\simeq 2hk_\fermi d$ for a small spin splitting $h$, and this is what produces the 0 to $\pi$ transitions \cite{buzdin_rmp_05}. For more complicated barriers, lacking inversion symmetry in addition to time-reversal symmetry, this same phenomenon can give rise to the $\varphi_0$ effect \cite{amundsen_rmp_24}, in which the current-phase relationship has a finite shift, $I(\varphi)\sim\sin(\varphi+\varphi_0)$. 

An altermagnet operates under the same principles, but in this case the spin splitting depends on the direction of the electron trajectory, as represented by its angle of incidence $\theta$ with respect to the interface. In addition, the effective distance traversed by current-carrying excitations across the barrier depends on their angle of propagation $\theta$ relative to the interface normal. Only particles travelling perpendicularly to the interface will travel a length $d$, whereas other particles travel a distance $d/\cos\theta$. This means that different trajectories generally will provide a different magnitude for the spin-dependent phase-accumulation $\Delta\psi$, even in the case of a ferromagnet with an isotropic spin-splitting in momentum space.  For a $d_{x^2-y^2}$ altermagnet with interface $\parallel \hat{y}$, the spin-splitting is maximal along the direction perpendicular to the interface, $\hat{x}$. For particles moving in trajectories at an angle, the length of the trajectory increases, but the spin-splitting of the altermagnet decreases. Therefore, $\Delta\psi$  actually decreases as $\theta$ increases. In contrast, for a $d_{xy}$ altermagnet, $\Delta\psi$ is zero for normal incidence $\theta=0$, but increases strongly when moving away from normal incidence since both the effective length and the spin-splitting felt by the particles increase. Since it is the magnitude of $\Delta\psi$ that determines how quickly the 0-$\pi$ transitions occur, it follows that for a wide junction the $0-\pi$ transitions occur more rapidly as a function of the junction length or altermagnetic spin-splitting in the $d_{xy}$-case compared to the $d_{x^2-y^2}$ case \cite{lu_prl_24}. In contrast, for a narrow junction where primarily modes close to normal incidence contribute, the situation is reversed \cite{ouassou_prl_23}.

\cref{sec:analytical} contains an analytical derivation of the effect that spin splitting has on the free energy of an altermagnetic Josephson junction, in a tunneling Hamiltonian approach summed to infinite order. We find the following analytical expression for the accumulated spin-dependent phase:
\begin{align}
\psi_s(\theta) = -4st_0 k_\fermi d\left[\cos 2\gamma\sin\theta + \sin 2\gamma\frac{\cos2\theta}{2\cos\theta}\right],
\end{align}
where $\gamma$ indicates the orientation of the altermagnet, with $\gamma = 0$ and $\gamma = \pi/4$ corresponding to the $d_{xy}$ and $d_{x^2-y^2}$ configurations, respectively, $s$ is the spin orientation, $k_F$ is the normal-state Fermi momentum and $t_0$ is the strength of the altermagnetism which enters the Hamiltonian in the form $\sim t_0[2k_xk_y - (k_x^2-k_y^2)]$.  A non-zero spin-splitting produces 0 to $\pi$ transitions. This spin splitting, in turn, depends on angle, and the total free energy is found by integrating over all angles. Nevertheless, we can find a rough estimate for the distance between transitions by employing the Gauss-Legendre quadrature, from which we can find an effective, integrated spin splitting by replacing $\sin\theta$ with $1/\sqrt{3}$, leading to
\begin{align*}
\Delta\psi = \frac{4}{\sqrt{6}}\left[2\sqrt{2}\cos2\gamma + \sin2\gamma\right]t_0 k_\fermi d.
\end{align*}
This means that the $d_{xy}$ altermagnet affects the junction with an effective spin splitting that is larger than that of $d_{x^2-y^2}$ by a factor of $2\sqrt{2}$, and  thus in the case of a wide junction produces 0 to $\pi$ transitions that are correspondingly more frequent as $t_0$ or $d$ is increased.

Another interesting aspect of 0 to $\pi$ transitions in Josephson junctions is how they enable the $\phi$ state. For general barrier strengths, the free energy is typically dominated by the lowest-order contribution in the tunneling Hamiltonian. As such, we can get a good estimate for the 0 to $\pi$ transitions by establishing where this contribution has a node. However, there are also higher harmonics present ($\cos n\varphi$, $n > 1$), and these terms do not generally vanish at the same transition points as the first harmonic. In fact, they behave as if the barrier length is multiplied by $n$, $d\to nd$. This means that in regions close to transition points, where the $n=1$ harmonic vanishes, the $n=2$ harmonic might dominate, which is a necessary, but not sufficient, condition \cite{goldobin_prb_07} for the $\phi$ state. 
In practice, the occurrence of a $\phi$-state in an altermagnet is rare, requiring a fine-tuned relation between $t_0$ and the junction length $d$. According to our numerical estimates, if a $\phi$-state occurs at a length $d=d^*$ at a fixed $t_0$, then only a slight deviation away from $d^*$ is sufficient to destroy the $\phi$-state \cite{beenakker_prb_23}.  However, we will in the following section show that the $\phi$-state is more resilient toward changes in the altermagnetic strength $t_0$ at a fixed junction length $d$.

\subsection{Numerical solution: generalized BTK-approach}

Having established an understanding of why a $\phi$-junction may appear in altermagnets, we turn to a purely numerical description for the purpose of achieving as realistic a modelling of such a system as possible. To this end, we make use of a generalized Blonder-Tinkham-Klapwijk (BTK) methodology \cite{blonder_prb_82} which is known to compare favorably to experimental measurements of transport in superconducting junctions \cite{wei_prl_98, jansen_natcom_24}. The basic idea in this approach is to set up a system of wavefunctions in Nambu-space that describe current-carrying excitations propagating in the system. In a JJ, these excitations are electron-like and hole-like quasiparticles in the superconductors whereas they are pure electrons and holes in the region separating the superconductors. By solving the Bogoliubov-de Gennes \cite{degennes_book_66, bogoliubov_jetp_58} equations in all regions of the system, one obtains the wavefunctions describing these excitations. Using microscopically derived boundary conditions, the wavefunctions and their derivatives are matched at the interfaces \cite{sun_prb_23, papaj_prb_23}. These boundary conditions can be summarized as a system of homogeneous linear equations on matrix form $A\boldsymbol{v}=0$, where the matrix $A=A(E)$ is a function of the energy $E$ carried by the current-carrying excitations while $\boldsymbol{v}$ is a vector containing the scattering coefficients that accompany each wavefunction describing an excitation. To ensure a non-trivial solution, we then proceed to solve the equation det$(A)=0$, which gives us a precise mathematical condition on which energies $E$ are allowed in the system. This energy is precisely that of the Andreev-bound states (ABS) carrying supercurrent through the non-superconducting region of the JJ. The permitted values for these bound states in an altermagnetic JJ will depend sensitively on both the superconducting phase difference $\varphi$ and the altermagnetic strength and band-structure orientation relative to the interfaces.

With the Andreev-bound state energies in hand, one can compute the zero-temperature free energy of the system in the short junction limit $d\ll \xi$ with $\xi$ the superconducting coherence length as
\begin{align}
    F = \sum_{k_y, E<0} E, 
    \end{align}
 Here, we aim to make close contact with an experimental setting and therefore incorporate realistic features such as a finite interface transparency and finite width and length of the junction, and we then compute the exact current-phase relation due to supercurrent-carrying Andreev-bound states. With regard to the altermagnetic parameters, we explore a wide range of strengths for both $d_{xy}$ and $d_{x^2-y^2}$ altermagnets  in order to provide experimental guidance for which system design provides the optimal qubit performance.

\begin{figure}
    \centering
    \includegraphics[]{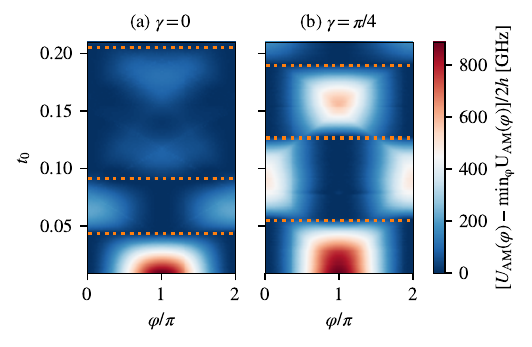}
    \caption{Altermagnetic Josephson potential $U_{\rm AM}(\varphi)$ as a function of the superconducting phase $\varphi$ and the altermagnetic strength $t_0$ for (a) pure $d_{xy}$ altermagnetism and (b) pure $d_{x^2-y^2}$ altermagnetism. The orange dashed lines are visual guides for the $0-\pi$ transitions upon increasing  $t_0$. We have subtracted the minimum value of $U_\text{AM}$. The parameters used are the junction length $d=0.2\xi$ and width $W=10d$ with $\xi$ the superconducting coherence length, the interface parameter $Z=0.4$, the superconducting gap $\Delta=\qty{0.15}{\milli \eV}$, and the Fermi wave vector $k_F\xi=100$. 
    } 
\label{fig:potential}
\end{figure}

\begin{figure}
    \centering
    \includegraphics{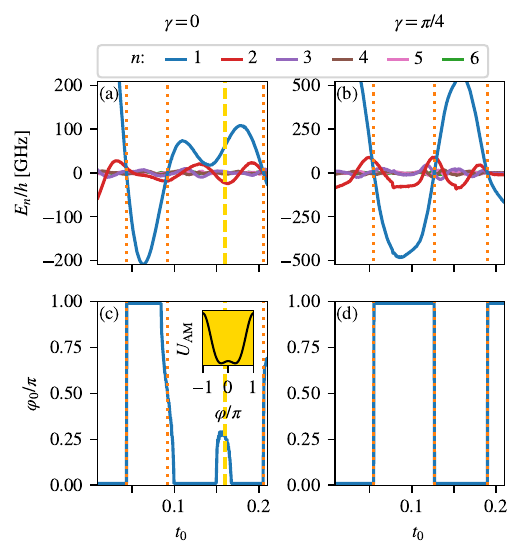}
    \caption{ Properties of the altermagnetic potential for $\gamma=0$ (left column) and $\gamma=\pi/4$ (right column). (a) and (b) show the harmonic content of the potential $U_{\rm AM}(\varphi) = - \sum_n E_n \cos(n\varphi)$. The coefficients $E_n$ were found using the method of least squares on the raw data, and the figure shows the first three components. (c) and (d) shows the phase value $\varphi_0$ at the minima of $U_{\rm AM}$ for $\varphi\in[0,\pi ]$. The inset in (c) shows the potential profile at $t_0 \approx  0.16$, and this $t_0$ strength  is marked with yellow dashed lines in (a) and (c). }
    \label{fig:harmonics}
\end{figure}

The altermagnetic Josephson potential $U_\text{AM}$ and its dependence on the superconducting phase difference $\varphi$ and altermagnetic strength $t_0$ are shown in Fig. \ref{fig:potential} for both $\gamma=0$ and $\gamma=\pi/4$. We have chosen an experimentally relevant set of parameters which places the system in the short-junction limit, where it is known that the phase-dependence of the free energy is primarily determined by ABS \cite{beenakker_prl_91}. The energy-phase relation is given by
\begin{equation}\label{eq:potentialcos}
    U_{\rm AM}(\varphi)=-\sum_{n \geq 1} E_n \cos(n\varphi),
\end{equation}
where the magnitude of the harmonic coefficients $E_n$ typically decrease with $n$. Any $\sin(n\varphi)$-terms are forbidden by symmetry in the present system \cite{lu_prl_24}. The numerical method gives noise in the free energy when it misses an ABS energy solution, and this numerical noise can only be described by including many higher harmonics. The first six harmonic coefficients of $U_{\rm AM}$ are shown in Fig. \ref{fig:harmonics}(a)-(b), and we observe that $E_1$ and $E_2$ dominate the energy-phase relation. Therefore, we truncate the sum at $n=2$ when calculating the qubit properties to capture the effects of the AM junction while avoiding features arising from numerical noise. When $E_1$ changes sign, a $0-\pi$ transition occurs. Clear $0-\pi$ oscillations are observed for both the altermagnetic orientations, with transitions occurring at the orange dashed lines.  The phase value $\varphi_0$ at the potential minimum is shown in Fig. \ref{fig:harmonics}(c)-(d). The $\pi$-state occurs frequently over a range of parameters and thus allows altermagnets to be used as stray-field-free $\pi$-shifters. In addition, we have searched for the appearance of a $\phi$-state where $U_\text{AM}$ displays a double-well potential with two degenerate minima at $\varphi=\pm\varphi_0$ which are not 0 or $\pi$. This occurs close to the $0-\pi$ transitions, as well as close to $t_0=0.16$ when $\gamma=0$ [see inset in Fig. \ref{fig:harmonics}(c)], where the first harmonic $E_1$ is small, allowing the higher harmonics to dominate the energy-phase relation. 

There are two ways that higher harmonics can become prominent in the energy-phase relation upon varying $d$: $0-\pi$ transitions and resonant tunneling.
First, the precise value of $t_0$ where the $0-\pi$ transitions and the $\phi$-state in Fig. \ref{fig:potential}(a) occur depends on the length of the junction $d$. This is because the aforementioned phase accumulation $\Delta\psi$ reaches resonant values $n\pi$ at different values of $t_0$ depending on $d$. Secondly, an additional effect of the finite length $d$ of the junction is that resonant tunneling may occur. This corresponds to certain lengths $d$ where the effective normal-state transmission coefficient is maximized due to constructive interference between right- and left-going excitations propagating through the non-superconducting region, which in turn affects the ABS-energy dispersion \cite{beenakker_prl_91}. When the condition for normal-state resonant tunneling is fulfilled, higher harmonics are amplified due to the improved transparency of the junction, despite the presence of barriers at the interfaces.

\begin{figure}
    \centering
    \includegraphics{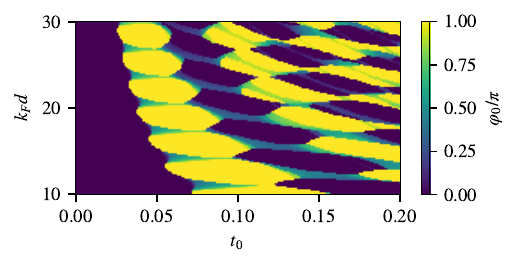}
    \caption{The location of the Josephson potential minima depends on the length $d$ and $t_0$ in the $\gamma=0$ configuration. When $\varphi_0\neq \{ 0, \pi \}$, the junction is in a $\phi$-state. We used the analytical expressions in \cite{beenakker_prb_23} with transmission probability $\Gamma=0.86$.}
    \label{fig:phistates}
\end{figure}

The $\phi$-state shows up in our numerical simulations in a narrow range close to $t_0=0.16$ in the case $\gamma=0$ as well as in the $0-\pi$ transitions close to $t_0=0.1$ and $t_0=0.2$. We find no $\phi$-state for $\gamma=\pi/4$, in accordance with analytical predictions \cite{beenakker_prb_23}. The free energy when the system is tuned precisely to the $\phi$-state is shown in the inset of Fig. \ref{fig:harmonics}(c) together with a plot of the contribution from different harmonics in Fig. \ref{fig:harmonics}(a) to the energy-phase relation. The latter verifies that the second harmonic is negative and sufficiently large compared to the first harmonic in order to push the minima of the Josephson potential to $\varphi_0 \neq \{0,\pi\}$ \cite{goldobin_prb_07}. We have verified that the $\phi$-state seen in our simulations also appears for the same parameter regime using the analytical results of \cite{beenakker_prb_23}. Moreover, we have used these analytical results to confirm that the $\phi$-state appears over a large variation of junction lengths and barrier transparencies. This is a very useful feature since, as we will show, the qubit performance is strongly enhanced in the $\phi$-state regime. The length and $t_0$ dependencies of $\varphi_0$ are shown in Fig. \ref{fig:phistates}.

\section{Altermagnetic qubits}\label{sec:singlequbit}

To assess how the altermagnetic JJ impacts qubit properties, we note that there are three archetypes of superconducting qubit designs: i) the transmon, ii) the flux qubit, and iii) fluxonium. For completeness, we list the relevant Hamiltonians describing the electrodynamics of each corresponding circuit below and comment on their typical main strengths and weaknesses. Starting with the transmon, its circuit Hamiltonian is:
\begin{align}
    H = 4E_c(\hat{n}-n_g)^2 + U_\text{AM}(\varphi)
\end{align}
We use the phase eigenstates as a basis in which case $\hat{n} = -i\partial_\varphi$ and $E_c = e^2/2C$ where $C$ is the capacitance of the circuit.  The transmon operates in the regime $|E_1|\gg E_c$ and thus gives a well-localized phase. While the protection against charge noise is very good, the anharmonicity is typically small since the eigenstates mostly probe a phase-regime where the potential is close to a harmonic oscillator. 

In the case of a flux qubit with one altermagnetic JJ and one $0$-junction with Josephson energy $E_J$, the circuit Hamiltonian instead takes the form:
\begin{align}
    H &= 4E_c(\hat{n}-n_g)^2 - E_J\cos(\varphi - 2\pi \Phi/\Phi_0) + U_\text{AM}(\varphi)\notag\\
    &+ \frac{(\Phi-\Phi_\text{ext})^2}{2L},
\end{align}
where $\Phi_{\rm ext}$ is the external flux through the loop, and $\Phi$ is the total flux through the loop.
The phase of the $E_J$ term follows from the flux quantization condition of the circuit. In the limit of small inductance $L$, the last term forces $\Phi \to \Phi_\text{ext}$. This allows the last term to be dropped if $\Phi$ is replaced with $\Phi_\text{ext}$ in the $E_J$-term. The flux qubit typically has larger anharmonicity than the transmon, but is more sensitive to noise. Note that for two conventional 0-JJs, an external flux of $\Phi = \Phi_0/2$, in addition to higher harmonics, is needed to produce the characteristic double-well potential of the flux qubit. On the other hand, if one JJ is replaced with a $\pi$-JJ that has sufficiently strong higher harmonics \cite{yamashita_prl_05}, the double-well potential emerges even at $\Phi=0$. Finally, for fluxonium the  Hamiltonian reads:
\begin{align}
    H = 4E_c(\hat{n}-n_g)^2 + U_\text{AM}(\varphi)
    + \frac{1}{2}E_L (\varphi-2\pi\Phi/\Phi_0)^2
\end{align}
where $E_L = E_J/N$ is the inductive energy associated with the superinductance formed by  $N$ series-coupled Josephson junctions with Josephson energies $E_J$. The fluxonium combines the charge noise insensitivity of the transmon with a large anharmonicity. The charge noise insensitivity is obtained through a large $U_\text{AM}/E_C$ ratio. Coming back to the role of the altermagnet, it enables a similar advantage as in the flux qubit case, namely an intrinsic $\pi$-state which enables a double-well potential in fluxonium at $\Phi=0$. For small $E_L$, the wells become widely separated, leading to an exponentially small energy splitting between the symmetric and antisymmetric combinations of the ground states in each well. At the same time, the spacing to higher excited states, corresponding to intra-well excitations, remains large, resulting in a strongly anharmonic spectrum. 

We will make the transmon design our primary focus in this work, and we consider two crystallographic orientations of the altermagnet. The setup is shown in Fig. \ref{fig:transmon}. To make contact with an experimentally relevant scenario, we note that typical values for the Josephson energy in a transmon are tens of GHz. This is much smaller than the values we numerically obtain in Fig. \ref{fig:potential} for most values of $t_0$. Nevertheless, a key point in our analysis is that we will show how superior qubit properties appear near $0-\pi$ transitions and the $\phi$-state caused by the altermagnetism. Near these transitions, the Josephson energy is strongly suppressed and in fact reaches levels of precisely tens of GHz, as seen in Fig. \ref{fig:harmonics}(a) and (b). Therefore, we choose an $E_c$ value of 0.25 GHz: for $t_0$ values away from the 0-$\pi$ transition points, the $E_J/E_c$ ratio is much larger than in typical transmons, but close to the transitions where we will focus our analysis for qubit operation, the ratio is standard \cite{krantz_apr_19}. Alternatively, should it be desirable to use the transmon with an AM JJ in a larger $t_0$ regime away from the transitions, $U_\text{AM}$ could be cleanly scaled down in amplitude without changing the current-phase relation simply by using a smaller value for the superconducting gap $\Delta$. In practice, this can be achieved by proximitizing for instance a semiconducting material via a host Al-superconductor, so that the proximitized region features a gap that is a fraction of $\Delta$ in the host superconductor.

\subsection{Qubit splitting and anharmonicity}
\begin{figure}
    \centering
    \includegraphics{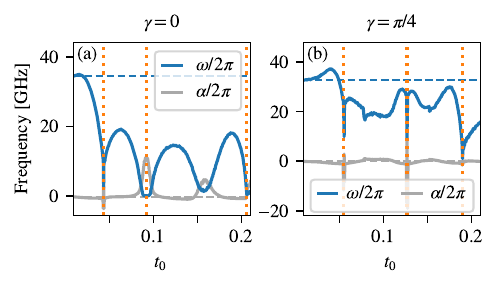}
    \caption{
    Qubit frequency $\omega$ (blue) and anharmonicity $\alpha$ (gray) for (a) $\gamma=0$ and (b) $\gamma=\pi/4$. The horizontal dashed lines represent the $t_0 \rightarrow 0$ values, while the orange dotted lines are visual guides for the $0-\pi$ transitions in the altermagnetic potential. We set $E_c/h=\qty{0.25}{\giga\hertz}$, which is a typical experimental value \cite{krantz_apr_19}.
    }
\label{fig:frequency_and_anharmonicity}
\end{figure}

The qubit splitting and anharmonicity for the altermagnetic transmon are shown in Fig. \ref{fig:frequency_and_anharmonicity} as functions of the altermagnet strength $t_0$. The anharmonicity is defined as $\alpha=\omega_{21}-\omega$, where $\omega_{21}=(E_2-E_1)/\hbar$ and $E_i$ are the eigenvalues. The results are shown for both crystallographic orientations $\gamma=0$ and $\gamma=\pi/4$. As a check of how the altermagnetism changes the qubit properties, we compare against a conventional SNS junction (dashed line), corresponding to the limit $t_0\to 0$. The altermagnetism generally lowers the qubit splitting across the entire $t_0$-range considered, particularly close to $0-\pi$ transition points. The qubit splitting lies in the range of a few GHz to 10 GHz close to the $0-\pi$ transitions, which falls squarely within typical experimental values for a transmon qubit \cite{roth_ieee_22}. Exactly at the $0-\pi$ transitions, the qubit splitting falls to sub-GHz due to the vanishing first harmonic in the Josephson potential, which leads to unfeasible temperature requirements for operating the qubit. However, as we will elaborate on below, this is not a problem since the ideal operational regime for an altermagnetic qubit is not right at the transition, but close to it.
At the same locations, the anharmonicity increases manyfold. This behavior can be understood from the fact that $U_\text{AM}$ is strongly suppressed near the $0-\pi$ transitions and the plasma frequency collapses as the accompanying Josephson energy shrinks. The enhancement of the anharmonicity arises from the competition between different harmonics in the Josephson energy. Near the transition, the first harmonic is strongly suppressed, while the second harmonic remains finite. This makes the quartic term larger relative to the quadratic term, and thus enhances the energy level anharmonicity, which can be shown as follows. Including the first and second harmonics in the potential yields $U_\text{AM}(\varphi) = -E_1\cos\varphi - E_2\cos(2\varphi).$ Expanding near $\varphi=0$, we obtain a ratio between the coefficients in front of the quartic and quadratic $\varphi$-terms of the form:
\begin{align}
    \frac{(E_1+16E_2)}{12(E_1+4E_2)}
\end{align}
If $E_2=0$, the ratio is $\frac{1}{12}$. If $E_1=0$, on the other hand, the ratio is much larger: $\frac{1}{3}$, which enhances the anharmonicity.

\begin{figure}
    \centering
    \includegraphics{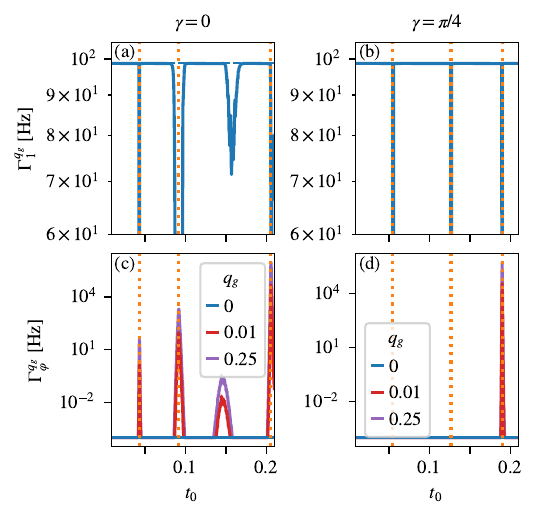}
    \caption{Decoherence in the altermagnetic transmon from charge noise for $\gamma=0$ (left column) and $\gamma=\pi/4$ (right column). (a) and (b) show the relaxation rate with $n_g=0$, and (c) and (d) show the dephasing rate at different $n_g$. Due to the large $U_\text{AM}/E_c$ ratio away from the $0-\pi$ transitions, which complicates an exact numerical solution, we have determined an upper limit (blue line) for the dephasing rate which in practice corresponds to a negligible $\Gamma_\varphi^{q_g}$.}
\label{fig:chargenoise}
\end{figure}

\subsection{Charge and dielectric noise sensitivity}

We here analyze the qubit response to charge and dielectric noise. 
The relaxation rate from charge noise is given by Fermi's golden rule, 
\begin{equation}\label{eq:relaxationrate}
    \Gamma_1^{q_g}=\frac{1}{\hbar^2} |\bra{0} 8E_c\frac{(\hat{n}-n_g)}{2e} \ket{1}|^2 S_{q_g}^{1/f}(\omega),
\end{equation}
where $S_{q_g}^{1/f}(\omega)$ is the charge noise spectral density evaluated at the qubit frequency. Experimentally, it is seen that charge noise typically has a $1/f$ spectral density:
\begin{equation}\label{eq:dephasingrate}
    S_{q_g}^{1/f}(\omega)=\frac{2\pi A_{q_g}^2}{|\omega|^\gamma}
\end{equation}
with $\gamma \approx 1$ and where $A_{q_g}=10^{ -4}e$ \cite{zorin1996background} is the noise amplitude. The dephasing rate from $1/f$ charge noise is given by \cite{groszkowski2018coherence}
\begin{equation}
    \Gamma_\varphi^{q_g}  =\sqrt{2A_{q_g}^2(\partial_{q_g} \omega)^2 | \text{ln}(\omega_{\rm ir}t)|},
\end{equation}
 where $\omega_{\mathrm{ir}}$ is an infrared cutoff set by the lowest frequencies present in the noise spectrum and thus the total effective measurement time~\cite{ithier_prb_05,paladino_rmp_14}. 
Experiments on superconducting qubits have observed $1/f$-type noise down to frequencies of order $10\,\mathrm{Hz}$~\cite{rower_prl_23}. Moreover, $t$ is the time during which the qubit is exposed to noise over a single cycle (shot) and accumulates phase fluctuations.
A typical experimental value for  $t$ is of order 10 $\mu$s \cite{krantz_apr_19}, which yields $\omega_{\mathrm{ir}} t \sim 10^{-4}$, corresponding to a logarithmic factor $\ln(1/\omega_{\mathrm{ir}} t)$ of order 10. 
We therefore adopt $\omega_{\mathrm{ir}} t = 2\pi \times 10^{-5}$ as a representative value. As a benchmark to compare our superconducting qubit  performance against, we note that state-of-the-art decoherence rates (including relaxation $\Gamma_1$ and dephasing $\Gamma_\varphi$) lie in the range $10^3-10^4$ Hz \cite{somoroff_prl_23, bland_nature_25}, with small $\Gamma$-values being desirable.

The decoherence rates for the charge noise in the altermagnetic transmon are shown in Fig. \ref{fig:chargenoise}. At the $0-\pi$ transition points (orange dashed lines), Fig. \ref{fig:frequency_and_anharmonicity} revealed that the qubit splitting becomes suppressed. This suggests that the noise spectral density $S_\lambda$ should become large due to its $1/|\omega|$ dependence, resulting in large decoherence rates.  However, we see that the opposite behavior occurs: the relaxation rate $\Gamma_{1}$ is suppressed near the $0-\pi$ transitions. This can be understood as a result of the dominance of higher harmonics in the current-phase relation. Near the 0--$\pi$ transition, the Josephson potential is dominated by a second harmonic $U_\text{AM}(\varphi) \approx -E_2 \cos(2\varphi)$, which is invariant under the discrete transformation $\varphi \rightarrow \varphi + \pi$. This symmetry leads to two equivalent potential wells at $\varphi = 0$ and $\varphi = \pi$, and the low-energy eigenstates take the form of symmetric and antisymmetric superpositions \cite{kitaev_arxiv_06, brooks_pra_13},
\begin{equation}\label{eq:doublewell wavefunctions}
    |0\rangle \simeq \frac{1}{\sqrt{2}}(\psi_L + \psi_R),\; 
|1\rangle \simeq \frac{1}{\sqrt{2}}(\psi_L - \psi_R),
\end{equation}
where $\psi_{L,R}$ are wavefunctions localized in the individual wells. This structure has direct consequences for relaxation due to charge coupling. The relaxation rate is governed by the transition matrix element of the charge operator $\Gamma_1 \propto |\langle 0|\hat{n}|1\rangle|^2.$
Evaluating this matrix element in the symmetric basis yields contributions from the two wells that enter with opposite sign,
\[
\langle 0|\hat{n}|1\rangle \sim 
\langle \psi_L|\hat{n}|\psi_L\rangle
- \langle \psi_R|\hat{n}|\psi_R\rangle
+ \text{interference terms}.
\]
The diagonal terms are strongly suppressed individually due to symmetry, whereas the interference terms are exponentially suppressed due to the tunneling barrier between the wells, as shown in \cite{dempster_prb_14}. As a result, the total matrix element is strongly suppressed, leading to a reduction of $\Gamma_1$ compared to when $U_\text{AM}(\varphi)$ features a single well. In the deep transmon limit that is realized away from the $0-\pi$ transition points, the dipole element $\langle 0|\hat{n}|1\rangle \propto (E_1/E_C)^{1/4}$ in the harmonic oscillator approximation \cite{koch_pra_07}.

However, there is another feature that stands out in Fig. \ref{fig:chargenoise}(a) that is not associated with any 0-$\pi$ transition. For $\gamma=0$, there is a clear drop in the relaxation rate close to $t_0=0.16$, even though there is no $0-\pi$ transition there. The origin of this feature is the $\phi$-state which emerges near this value of $t_0$. The double-well structure ensures that the qubit states, which have opposite parity, also have exponentially suppressed overlap in the barrier region, rendering $\langle 0|\hat{n}|1\rangle$ strongly suppressed for the same reason as described above. This shows that the $\phi$-state can lead to similar qubit properties as when the system is close to the $0-\pi$ transition, although the sign of $E_1$ never changes and thus the system never realizes an actual transition.

Turning to the dephasing plots, these are numerically substantially more demanding to compute than the relaxation rates, as a much higher number of states is required in the phase basis, and it is convenient to diagonalize the Hamiltonian in the charge basis instead. The point $n_g=0$ is a sweet spot for dephasing in our model (see Appendix \ref{sec:appendixdephasing}), but to avoid an idealized scenario we instead compute dephasing at finite values of $n_g$. The dephasing rate in our setup becomes extremely small far away from the $0-\pi$ transition points where $U_\text{AM}/E_c \sim 10^2 - 10^3$, which is larger than typical experimental values. However, our main regime of interest is close to the $0-\pi$ transitions and the $\phi$-state where the ratio of the Josephson energy to $E_c$ is $\sim 50$, which is a typical experimental value. Near these points, the second harmonic is stronger relative to the first harmonic compared to elsewhere, but still small in magnitude. Therefore, the dephasing due to charge noise increases near these points in our setup. In practice, the actual value of the dephasing nevertheless does not exceed $10^3$ Hz, indicating lifetimes $> 1$ ms, which is an excellent qubit property. Thus, the increased dephasing near $0-\pi$ transitions in our setup still remains small in magnitude with one exception: the last $0-\pi$ transition for both $\gamma=0$ and $\gamma=\pi/4$ (respectively $t_0 \simeq 0.20$ and $t_0\simeq 0.18$). The reason for this is that at this particular transition, both the first and second harmonic vanish, thus causing the qubit states to be essentially plane waves, which are extremely sensitive to $n_g$ due to a large degree of charge-number localization.

\begin{figure}
    \centering
    \includegraphics[]{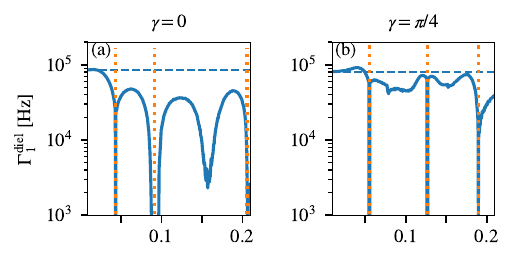}
    \caption{Relaxation rate for dielectric noise in the altermagnetic transmon in the (a) $\gamma=0$ configuration and (b) $\gamma=\pi/4$ configuration. The horizontal dashed lines represent the $t_0 \rightarrow 0$ values, while the orange dotted lines are visual guides for the $0-\pi$ transitions in the altermagnetic potential. The smallest relaxation rates are found in the proposed operational regime consisting of the vicinity of $0-\pi$ transitions and the $\phi$-state. 
    }
    \label{fig:dielectric}
\end{figure}

A noise channel that severely limits the lifetime of transmons is dielectric loss. Transmons are typically grown on a substrate that contains defects which can be modelled in an approximate manner as two-level quantum systems that emit and absorb photons through electromagnetic fluctuations. When this happens, it can cause current fluctuations through the Josephson element of the qubit, as charges are excited and de-excited. In other words, dielectric noise leads to current
noise. While dephasing from dielectric noise can be completely disregarded in our setup for zero current bias (see Appendix \ref{sec:appendixdephasing}), relaxation from dielectric noise is calculated from
\begin{equation}
    \Gamma_1^{\rm diel} = \hbar |\bra{0}\varphi\ket{1}|^2\frac{\omega^2 \tan\delta_{\rm diel}}{4E_c}\left[\coth\left(\frac{\hbar\omega}{k_BT}\right)+1\right], 
\end{equation}
where $\tan \delta_{\rm diel} = 1/Q_{\rm diel}$ is the loss tangent extracted from the dielectric loss quality factor $Q_{\rm diel}$. An experimentally realistic value is of the order $\tan \delta_{\rm diel} =\qty{2e-7}{}$ \cite{krojer_prr_24}, and it can be reduced via substrate engineering \cite{bland_nature_25}.  When calculating the matrix element numerically, we note that the $\varphi$ interval of length $2\pi$ should be chosen such that the phase edges are located at the potential maxima, where the wave function is suppressed. This is to avoid numerical artifacts arising from abrupt jumps in $\varphi$ at the phase boundaries.  The relaxation rate for the altermagnetic transmon due to dielectric noise is shown in Fig. \ref{fig:dielectric}, with rates of the order $10^4-10^5$ Hz when the altermagnetic junction rests in a 0- or $\pi$-state. This is a bottleneck for the qubit lifetimes. In these regimes, we can neglect higher harmonics in the Josephson potential and approximate the potential as a harmonic oscillator. The relaxation rate then reduces to $ \Gamma_1^{\rm diel} = \sqrt{32E_1E_c}\tan \delta_{\rm diel}/\hbar$ at temperatures $k_B T\ll \hbar \omega$. The high relaxation rate can therefore be traced to the large magnitude of the free energy. If the qubit is intended to be operated in this regime, it might therefore be advantageous to reduce the free energy by swapping the superconductor with a material with a weaker proximity-induced superconducting gap. Importantly, in the regions where the higher harmonics dominate, i.e. near the $0-\pi$ transitions and the $\phi$-state in the $\gamma=0$ configuration, the relaxation rate drops drastically. In the limit $|E_1|\ll|E_2|$, the wave functions can again be approximated by Eq. \eqref{eq:doublewell wavefunctions}. The matrix element in Fermi's golden rule becomes equal to half the distance between the two wells: $\bra{0}\varphi\ket{1}=\pi/2$. The relaxation rate is thus governed by the qubit frequency as $\Gamma_1^{\rm diel} \sim 4\times 10 ^{-8}\frac{\omega^2}{E_c/h}$. At small frequencies where $k_BT \gg \hbar \omega$, we can approximate $\coth(\hbar\omega/k_BT)+1 \approx k_BT/\hbar\omega$, so $\Gamma_1^{\rm diel} \sim 2\times 10 ^{-8}\frac{\omega}{E_c/h}\frac{k_BT}{\hbar}$. At the $0-\pi$ transitions, the small qubit frequency reduces the relaxation rate from dielectric noise below $\qty{1e3}{\hertz}$. In the $\phi$-state, the matrix element can be approximated by $\varphi_0$, and the relaxation rate from dielectric noise is below $10^4$ Hz as seen at $t_0\approx 0.16$ in Fig. \ref{fig:dielectric}(a). Since these rates are more than an order of magnitude smaller than in the $0$- and $\pi$-states, this further strengthens our proposal of operating the qubit near the $0-\pi$ transitions and in the $\phi$-state.

\subsection{Flux qubit and fluxonium}
We now turn to the two other types of superconducting qubit designs: the flux qubit and fluxonium. The defining properties of the flux qubit are its double-well potential giving rise to persistent currents in its qubit states and increased anharmonicity. The difficulty in assessing the impact of altermagnetism in this scenario is that when sweeping the strength $t_0$ for a flux qubit including a regular 0-junction with Josephson energy $E_{J,0}$ and our altermagnetic JJ, we find that the double-well potential is realized only in a quite limited regime of $t_0$ values. In general, we find that it is realized when $U_\text{AM}(\varphi)$ (i) contains substantial higher harmonics and (ii) is not too large compared to $E_{J,0}$. These are similar conditions as in Ref. \cite{yamashita_prl_05}, and confirm that altermagnetic JJs can be used to create stray-field-free flux qubits operating at zero external flux. In the case of fluxonium, we find mostly similar behavior for the qubit performance parameters as in the transmon case, with characteristic behavior near 0-$\pi$ transitions, such as increased anharmonicity and reduced relaxation rate, and in a $\phi$-state. For this reason, we have shown numerical results only for the transmon architecture in this work. There is nevertheless one key difference between fluxonium and the transmon design: fluxonium changes its properties depending on whether $U_\text{AM}(\varphi)$ is in a 0 or $\pi$-state, whereas the transmon does not. A $\pi$-state junction embedded in a fluxonium loop produces a double-well potential at zero 
external flux, since the inverted cosine potential and the harmonic restoring force compete to 
create two symmetric minima at $\pm\varphi_0$. Consequently, for an altermagnetic junction that 
supports both $0-\pi$ transitions and an intrinsic $\phi$-state, the fluxonium geometry 
gives rise to double-well behavior across a larger portion of the $t_0$ parameter space. This occurs both
when the junction resides in its intrinsic $\phi$-state and when it resides in the $\pi$-state, with the sweet spot shifting from half flux ($\Phi_\text{ext}/\Phi_0 = 1/2$) to zero 
flux ($\Phi_\text{ext} = 0$) in the latter case. Note how the two mechanisms are physically distinct. In the $\phi$-state case the double well is inherited from the junction's intrinsic energy 
landscape and relative ratio of first and second harmonic in the energy-phase relation, whereas in the $\pi$-state case, it is generated by the interplay between the 
inductive term and the sign of $E_1$.

\section{Operation: gates, initialization, and readout}

\subsection{Single-qubit rotation}
Following initialization to the ground state via relaxation or readout, both state preparation and gate operations for the altermagnetic transmon are implemented using resonant microwave drives, which induce controlled rotations on the Bloch sphere. These processes are governed by the charge dipole matrix element, and are therefore equally sensitive to modifications of the current–phase relation introduced by altermagnetism. The microwave drive is incorporated into the Hamiltonian by an extra term 
\begin{align}\label{eq:drive}
    H_\text{drive}(t) = V_0\sin(\Omega t)\hat{n}
\end{align}
where $\Omega$ is the drive frequency and $V_0$ its amplitude.
\begin{figure}[h!]
\includegraphics{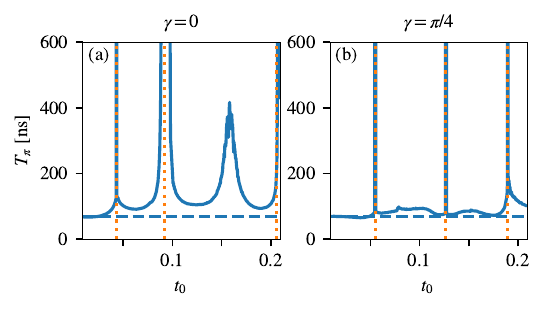}
\caption{The $\pi$-pulse time $T_\pi$ as a function of altermagnet strength $t_0$ in (a) the $\gamma=0$ configuration and (b) the $\gamma=\pi/4$ configuration. The horizontal dashed lines represent the $t_0 \rightarrow 0$ values, while the orange dotted lines are visual guides for the $0-\pi$ transitions. We have used $V_0/h=2.5$ MHz for the drive voltage. 
}
\label{fig:singleT}
\end{figure}
In \cref{fig:singleT}, we plot the time $T_{\pi}$ (half of a Rabi oscillation) required to induce transitions from $\ket{0}$ to $\ket{1}$ by using an AC voltage driving term $\sim \sin(\Omega t)$ at the resonance frequency $\Omega=\omega$ where $\omega$ is the qubit splitting. This is essentially population inversion taking place over the $\pi$-pulse time $T_\pi$. The gate time can be determined numerically by applying the time-evolution operator, including the drive term in the Hamiltonian, to the initial state $\ket{0}$ and then identifying the overlap of the quantum state with $\ket{1}$ at a given time. In practice, it is computationally a very good approximation to determine the gate time from the rotating-wave approximation (RWA)
expression \cite{krantz_apr_19}
\begin{align}\label{eq:analyticalT}
T_\pi = \frac{h}{2V_0 |\langle 0|\hat{n}|1\rangle|}
\end{align}
as long as the voltage term is very small compared to $E_c$ and $U_\text{AM}(\varphi)$ and driven at resonance (qubit splitting). We have numerically confirmed, by truncating Hilbert space to the seven lowest eigenstates and performing a full unitary time-evolution due to the drive voltage, that we obtain excellent quantitative agreement with the gate times based on Eq. (\ref{eq:analyticalT}), even very close to $0-\pi$ transitions and in the $\phi$-state. The resolution of clean Rabi oscillations when using the full time-evolution, which is beyond an effective two-level model and the RWA, confirms that matrix elements corresponding to virtual excitations give a negligible contribution to population inversion throughout the entire range of $t_0$. This is likely because of their small magnitude and the off-resonant nature of the driving for higher levels. \\

For $t_0$-values that are not in the vicinity of $0-\pi$ transitions, our choice of $V_0/h = \qty{2.5}{\mega \hertz}$ gives rise to gate times in the range 50-100 ns. Although this exceeds state-of-the-art values \cite{yaoyao_nsr_25}, it corresponds to a drive strength that gives weak leakage and is adequate for demonstrating proof-of-principle operation. The most striking feature is nevertheless the large peaks occurring at certain values of $t_0$, where the AC drive becomes inefficient at causing transitions and thus loses its function as a gate. The physical origin of these peaks can be identified as follows. The values of $t_0$ where the peaks arise are close to $0-\pi$ transitions, as seen by comparing with Fig. \ref{fig:potential}. At the $0-\pi$ transition, the second-harmonic potential produces a symmetric double-well structure, and the qubit states take the form of symmetric and antisymmetric superpositions of the same localized wavefunctions. Driving via an AC voltage couples to the charge operator according to Eq. (\ref{eq:drive}). Projected onto the qubit subspace, the transition amplitude is set by the transition matrix element $\langle 0|\hat{n}|1\rangle$. As discussed above, this matrix element is strongly suppressed near the transition due to destructive interference between contributions from the two wells. Consequently, the charge operator couples only weakly to the qubit, leading to a strongly reduced Rabi frequency and hence long gate times. We note that at the $0-\pi$ transition points, where the direct dipole matrix element is strongly suppressed, transitions via virtual excitations will play a bigger role in determining the gate time and may become the dominant mechanism for population transfer. Although these processes are suppressed in absolute magnitude for weak drive $V_0/E_c \ll 1$, they cannot be neglected once the leading-order coupling vanishes, and are expected to limit the growth of the gate time.

Comparing the figures showing the qubit anharmonicity and decoherence times with the gate operation time, one observes that the 0-$\pi$ oscillations present a fundamental trade-off. On the one hand, the anharmonicity and decoherence rates are optimized close to the $0-\pi$ transition points. On the other hand, the magnitude of the gate operation time $T_{\pi}$ is untenable for practical purposes at the precise location of these points. To solve this, we propose a protocol where the qubit idles near the $0-\pi$ transition since this is a protected regime with long lifetimes. During gate operations, the qubit is temporarily detuned away from this regime, and then returned. The key insight is that both the altermagnetic strength and the spin-dependent anisotropy in the band structure govern the spin-split band structure and hence the precise location of the $0-\pi$ crossing, and that this can be modified \textit{in situ} by applying strain to the altermagnetic layer. This can be achieved by depositing a thin altermagnetic film on a piezoelectric substrate. Applying a gate voltage induces a substrate expansion or contraction, and the mechanical deformation transfers across the interface to the altermagnet, modifying its lattice constants and spin-dependent hopping integrals. It was recently shown in \cite{khodas_prb_26} that the net effect of strain in an altermagnet is to rescale the two spin-split Fermi surfaces, and also induce a small net magnetization. More specifically, assuming an experimentally realistic strain magnitude that modulates the resulting hopping parameters by at most a few percent \cite{xu_natcom_22, khodas_prb_26}, such a magnitude is already sufficient to significantly mitigate the long gate times. Having moved away from the points where $T_{\pi}$ is very large, gate operations can now be performed via standard microwave voltage drive, after which strain is released and the qubit returns to the protected idle point. 

Piezoelectric strain applied to magnetic thin films is well-established experimentally \cite{geprags_ssc_14}. The proposed protocol requires no external magnetic field, and voltage-driven piezoelectric actuators can operate in the MHz range \cite{liu_nature_20}. However, there are experimental challenges as well. The piezoelectric substrate will introduce additional noise channels \cite{zhou_natcom_25} , and release of the strain could generate a transient phonon mode \cite{choi_arxiv_25}. This mode would have to relax before going back to the original idle qubit state, requiring an optimized substrate geometry to minimize such an effect. The ideal hierarchy of time scales for strain-control over the protected qubit regime would be
\begin{align}
    T_\text{strain} \ll T_{\pi} \ll \Gamma^{-1}
\end{align}
where $T_\text{strain}$ is the time it takes to modulate the strain. Gate times for a single operation in superconducting qubits typically lie in the range 10-100 ns, consistent with Fig. \ref{fig:singleT} except right at the $0-\pi$ transition points, so that a MHz-operation for the piezoelectric actuator can satisfy the above hierarchy if one performs several gate operations subsequently.

Should the disadvantages of using piezoelectric strain prior to gate operation prove too challenging, the best solution is to instead operate further away from the $0-\pi$ transition points, offering much better gate operation times at the expense of worse anharmonicity and decoherence.

\subsection{Readout}
The qubit can be read out via dispersive coupling to a microwave resonator, described at the microscopic level by an interaction of the form 
\begin{align}
H_{\mathrm{int}} = g_r\, \hat{O}(a + a^\dagger),
\end{align}
where $\hat{O}$ is the operator through which the qubit couples to the electromagnetic field. The resonator itself is described by the Hamiltonian $H_r=\hbar \omega_r a^\dagger a$. In the dispersive regime, where the transition frequencies of the entire transmon system are sufficiently far detuned from the cavity frequency, the qubit state induces a shift of the resonator frequency that can be extracted directly from the spectrum of the full coupled Hamiltonian. To do so, we identify the resonator transition frequencies conditioned on the qubit being in its ground or excited state,
\begin{align}
\hbar\omega^r_{0} = E_{0}^{(1)} - E_{0}^{(0)}, \quad
\hbar\omega^r_{1} = E_{1}^{(1)} - E_{1}^{(0)},
\end{align}
where \(E_{n}^{(k)}\) denotes the eigenenergy corresponding predominantly to the qubit state $n$ and photon number $k$. The dispersive shift is then defined operationally as
\begin{align}
\chi = \frac{1}{2}\left(\omega^r_{1} - \omega^r_{0}\right).
\end{align}
This definition does not rely on perturbation theory and remains valid beyond the regime where an effective two-level description in terms of Pauli operators is applicable. Instead, $\chi$ reflects the full hybridization between qubit and resonator degrees of freedom. The overlap factor is determined by the overlap between the wave functions of the dressed system, belonging to the eigenenergies $E_{0}^{(0)}, E_{0}^{(1)}, E_{1}^{(0)}$ and $E_{1}^{(1)}$, that have the largest overlap with the wavefunctions of the uncoupled system with the same qubit state and photon number.

When our altermagnetic transmon is capacitively coupled to the readout resonators, $\hat{O} = \hat{n}$, and the dispersive shift $\chi$ depends on the matrix element  $\langle 0 | \hat{n} | 1 \rangle$.
Near the 0–$\pi$ transition, the qubit eigenstates become strongly delocalized in phase space. As a result, the matrix element $\langle 0|\hat{n}|1\rangle$ is strongly suppressed, as we explained earlier. This does not necessarily mean that readout becomes a dark mode near these transitions, since the dispersive shift also depends on the detuning between the qubit and the resonator as well as virtual excitations. To determine the dispersive shift for readout of our altermagnetic transmon, we choose \(\omega_r/2\pi = 15\,\mathrm{GHz}\), close to demonstrated experimentally relevant microwave
readout-resonator frequencies for superconducting transmon circuits \cite{zotova_apl_24}. With regard to the
qubit--resonator coupling, experiments quote coupling strengths of order
$g_r/h = 100\,\mathrm{MHz}$ 
\cite{jeffrey_prl_14}. Finally, as an experimentally relevant readout
target, we take
$|\chi|/2\pi > 1\,\mathrm{MHz}
$ \cite{delia_appsci_24}, provided the dressed eigenstates retain high overlap with the
corresponding bare qubit--photon states.

\begin{figure}
    \centering
    \includegraphics{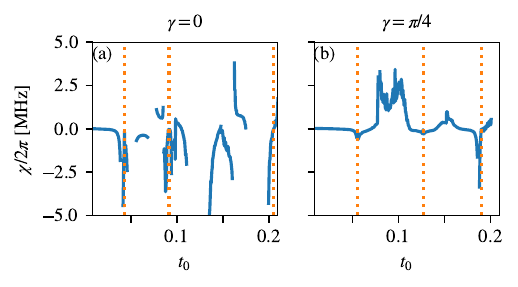}
    \caption{Dispersive shift for (a) the $\gamma=0$ configuration and (b) the $\gamma=\pi/4$ configuration. We only show the $t_0$-values for which the overlap factor exceeds $99$\%. The resonator frequency is set to  $\omega_r/2\pi = \qty{15}{\giga \hertz}$ and the qubit/resonator coupling strength is $g_r/h=\qty{100}{\mega\hertz}$.
    }
    \label{fig:readout}
\end{figure}

In Fig. \ref{fig:readout}, we plot the dispersive shift $\chi$ as a function of the altermagnetic strength $t_0$ for both crystallographic orientations $\gamma=0$ and $\gamma=\pi/4$. We plot $\chi$ only for $t_0$-values where the overlap between the dressed and bare states exceeds 99\%, as required for actual operation of the qubit. This requirement is satisfied in the dispersive limit when the detuning $\Delta =|\omega-\omega_r|\gg g_r$. We observe that this requirement is not met at the missing $t_0$ values in Fig. \ref{fig:readout} since they match the $t_0$ values for which the qubit frequency in Fig. \ref{fig:frequency_and_anharmonicity} are close to $\omega_r$. The dispersive shift is approximately given by $\chi\approx g_r^2/\Delta$ \cite{krantz_apr_19}, so it is small when the qubit frequency is large compared to $\omega_r$, such as for small $t_0$ values. Close to the $0-\pi$ transitions and the $\phi$-state, the detuning is smaller since we chose $\omega_r$ to give a proper detuning at these altermagnetic strengths. Moreover, the dispersive shift has a plateau at about $\chi/2\pi=2$ MHz in the $\gamma=\pi/4$ configuration close to $t_0=0.1$, as seen in Fig. \ref{fig:readout}(b). Here, the qubit frequency is $\omega/2\pi \approx 20$ GHz, which is an optimal regime for dispersive readout with a resonator frequency of $\omega_r/2\pi=15$ GHz: it strikes a balance between having a detuning $\Delta$ relative to $g_r$ that is small enough to give a large shift $\chi$, but large enough that the overlap between dressed and bare states is close to unity.

\subsection{Coupled altermagnetic qubits: CZ- and CNOT-gate}\label{sec:doublequbit}
To achieve universal quantum computation, a system must support arbitrary single-qubit operations together with at least one entangling two-qubit gate. A convenient choice is the controlled-phase (CZ) gate. The CZ gate involves phase-shifts for the qubit-states and is a natural gate in systems where interactions generate conditional phases \cite{Wendin2017}, precisely like our capacitively coupled, altermagnetic transmons.

The CZ gate acts diagonally in the computational basis, leaving $\ket{00}$, $\ket{01}$, and $\ket{10}$ unchanged while applying a minus sign to $\ket{11}$. Despite its simple form, it is an entangling operation, as it generates non-separable states when acting on superpositions. Its implementation relies on the accumulation of a conditional phase rather than population transfer. In a coupled two-qubit system, this conditional phase originates from the interaction-induced energy
\begin{equation}
\zeta_{ZZ} = E_{11} - E_{10} - E_{01} + E_{00},
\end{equation}
which removes additive single-qubit contributions and isolates the non-additive (interaction) part of the spectrum. In other words,  $\zeta_{ZZ}$ captures how the energy of one qubit depends on the state of the other, and quantifies the conditional phase accumulated between the two-qubit basis states after compensating for the independent phase evolution of each qubit. A CZ gate is obtained when this interaction-induced phase reaches $\pi$, corresponding to a gate time
\begin{equation}
t_{CZ} = \frac{\hbar\pi}{|\zeta_{ZZ}|}.
\end{equation}

Two key quantities characterize the performance of the CZ gate. First, the magnitude of $\zeta_{ZZ}$ sets the gate time, with larger values enabling faster entangling operations. Second, it is essential to verify that the eigenstates of the coupled system remain close to the computational (bare) basis states. This can be quantified via overlaps $|\langle \tilde{ab} | ab \rangle|^2$, where $|\tilde{ab}\rangle$,  $ab \in \{00,01,10,11\}$ are the dressed eigenstates. High overlap ensures that the qubit subspace is well defined and that leakage to non-computational states is negligible. The dressed computational states are identified by assigning each bare computational state to the eigenstate of the coupled Hamiltonian with maximum overlap, excluding eigenstates already assigned. This is a sequential maximum-overlap labeling procedure.

In the following, we evaluate these quantities by using the altermagnetic Josephson energy $U_\text{AM}(\varphi)$ as input to the transmon Hamiltonian, allowing us to extract both the interaction strength $\zeta_{ZZ}$ and the corresponding gate time, as well as the overlap with the computational basis. To ensure experimental relevance, we set a condition that the minimum   of the probability overlaps satisfy
\begin{equation}\label{eq:conditionQ}
\min_{ab \in \{00,01,10,11\}} |\langle \tilde{ab} | ab \rangle|^2 \equiv Q > 99\%,
\end{equation}
ensuring that the qubit subspace is well preserved and that leakage to non-computational states remains negligible.

\begin{figure}[h!]
\includegraphics[]{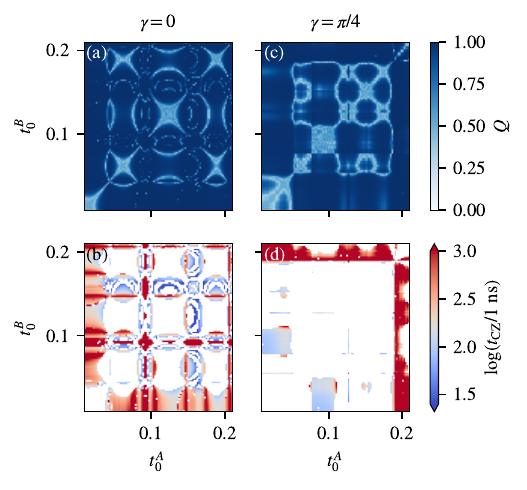}
\caption{CZ gate with coupling strength $g/h=100$ MHz and altermagnetic configurations $\gamma=0$ in (a)-(b)  and $\gamma=\pi/4$ in (c)-(d) for qubits A and B with altermagnetic strengths $t_0^A$ and $t_0^B$, respectively. Upper row: The minimal overlap $Q$ of the four lowest bare and dressed two-qubit states. Bottom row: CZ gate time. The color maximum represents gate times $t_{CZ}>\qty{1000}{\nano \second}$ and the color minimum represents $t_{CZ} \simeq 10$ ns.
}  
\label{fig:doublegatet1}
\end{figure}

\begin{figure}[h!]
\includegraphics{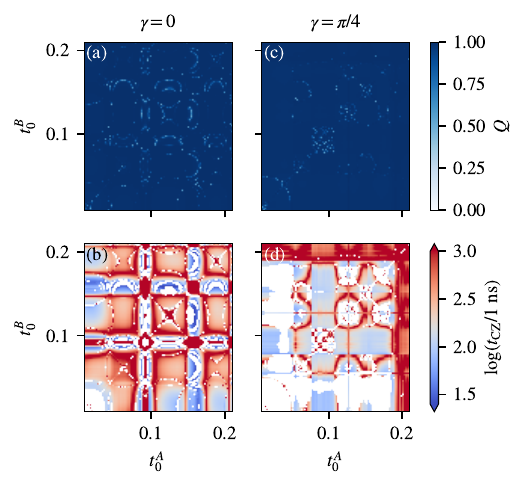}
\caption{CZ gate with tailored coupling strength $g=\hbar |\omega^A-\omega^B|/200$ MHz and altermagnetic configurations $\gamma=0$ in (a)-(b)  and $\gamma=\pi/4$ in (c)-(d) for qubits A and B with altermagnetic strengths $t_0^A$ and $t_0^B$, respectively. Upper row: The minimal overlap $Q$ of the four lowest bare and dressed two-qubit states. Bottom row: CZ gate time. The color maximum represents gate times $t_{CZ}>\qty{1000}{\nano \second}$ and the color minimum represents $t_{CZ} \simeq 10$ ns.
}
\label{fig:tailoredCZ}
\end{figure}

The results are shown in Fig. \ref{fig:doublegatet1} for a realistic coupling strength of $g/h=100$ MHz where the coupling term in the Hamiltonian is $g\hat{n}_1\hat{n}_2$. This is a larger coupling than typical always-on capacitive couplings in experiments \cite{howard_prr_23}, but matches values for tunable couplers \footnote{In this work, \(g\) denotes the bare coefficient of the charge-charge term
$H_{\mathrm{int}} = g\, n_1 n_2 .
$ In experiment, it is often the  transverse coupling \(J\) that is quoted, obtained after projecting the charge operators into the relevant low-energy subspace,
$
\frac{J}{h}
\simeq
\frac{g}{h}
\left|\langle 0 | n_1 | 1 \rangle\right|
\left|\langle 0 | n_2 | 1 \rangle\right| .$
For transmon-like states, $\left|\langle 0 | n | 1 \rangle\right|
\approx
\left(\frac{E_J}{32 E_C}\right)^{1/4},$
so \(J\) and \(g\) are of the same order but differ by the charge matrix elements.} during gate operation \cite{chen_prl_14}. After diagonalizing the individual qubit Hamiltonians, we truncate the Hilbert space of each qubit to the eight lowest-energy eigenstates, thus accounting for virtual excitations within the retained low-energy subspace. The two-qubit system is then constructed in the tensor-product space of these truncated bases, including the interaction term proportional to $g$. The resulting Hamiltonian is diagonalized to obtain the dressed eigenstates and corresponding eigenvalues of the coupled system. The overlap factor $Q$ in Fig. \ref{fig:doublegatet1}(a) is plotted against the altermagnetic strength $t_0$ in the two qubits (A and B) in the $\gamma=0$ crystallographic junction configuration, revealing that $Q\to 1$ is satisfied in a large parameter regime. The most notable exception is the diagonal line where $t_0^A = t_0^B$ where $Q \simeq 0.5$ for most of the line, which can be understood as follows. Along the symmetric line, the bare states $\ket{01}$ and $\ket{10}$ are degenerate in the absence of coupling. In this case, any linear combination of these states is an eigenstate. However, once a finite interaction term $g$ is included, the two states are generally coupled through matrix elements of the form $\langle 10| n_1 n_2 |01\rangle \neq 0$. As a result, the degeneracy is lifted, yielding symmetric and antisymmetric combinations $(\ket{01} \pm \ket{10})/\sqrt{2}$ as the true eigenstates, with a small energy splitting determined by the strength of the coupling. Importantly, this rotation of the eigenbasis occurs for arbitrarily small but nonzero coupling, meaning that $\ket{01}$ and $\ket{10}$ are no longer eigenstates even if their energies remain nearly identical. Consequently, the reduced overlap with the bare computational basis along the diagonal reflects a genuine hybridization of states due to the interaction, rather than a breakdown of the qubit description.

In contrast to a single qubit, where peaks in the gate time occur at the $0$–$\pi$ transition points, we find that the system behavior is more complex in the coupled qubit case. This reflects the fact that the interaction strength $\zeta_{ZZ}$ depends on the combined properties of both qubits, and cannot be understood solely in terms of single-qubit matrix element suppression. Nevertheless, Fig. \ref{fig:doublegatet1}(a) also demonstrates poor gate performance on the diagonal near the points where $t_0^A=t_0^B$ feature 0-$\pi$ transitions in the form of broad, star-like areas with low overlap factor $Q$. This reflects precisely the fact that near the transitions, the effective interaction becomes strong relative to the level splittings due to the suppression of $U_\text{AM}$, and the overlap between the dressed and bare states is reduced even though the relevant two-qubit subspace remains well defined.
In Fig. \ref{fig:doublegatet1}(b), the gate time $t_{CZ}$ is shown in log-scale. To clearly visualize only gate times of practical relevance where Eq. (\ref{eq:conditionQ}) is fulfilled, we have omitted from the plot (shown as white) any regions where $Q<99\%$. Even with this condition, there is a considerable area in the $t_0^A-t_0^B$ plane where fast gate times can be achieved. Typical experimental values for $t_{CZ}$ in capacitively coupled transmon qubits are $\sim$ 200 ns \cite{caldwell_pra_18}, corresponding to a log-value of 2. We see that even smaller times can be reached at selected regions in Fig. \ref{fig:doublegatet1}(b). Since we propose to operate the altermagnetic qubit close to the $0-\pi$ transitions, it would be advantageous to have fast gate times at such $t_0$-values. It is seen from Fig. \ref{fig:doublegatet1} that while this does not occur along the diagonal, there are indeed limited regimes where both qubits are near distinct $0-\pi$ transition points, such as $t_0^A \simeq 0.05$ and $t_0^B \simeq 0.09$, where the gate times are excellent. Interestingly, if both $t_0^{A,B}$ are close to $0.16$, such that both junctions are in the $\phi$-state, the gate times are low, revealing an additional strength of utilizing the altermagnetic $\phi$-state for improved qubit functionality. The results for the $\gamma=\pi/4$ orientation shown in Fig. \ref{fig:doublegatet1}(c) and (d) display quantitative differences from (a) and (b) such as generally smaller overlap between the dressed and bare states, restricting the operational regime for the CZ gate, as well as larger gate times. Otherwise,  Fig. \ref{fig:doublegatet1}(c) and (d) feature similar behavior near distinct $0-\pi$ transition points for the two qubits in terms of limited regions of fast gate times. The dependence of $\zeta_{ZZ}$ on the altermagnetic potential thus provides a direct link between the energy–phase relation  of the altermagnetic JJs and the corresponding two-qubit gate performance.

The above analysis of the CZ-gate at a fixed coupling strength $g$ provides a useful baseline for assessing the intrinsic dependence of the gate performance on the altermagnetic strengths of the two qubits. In this fixed-coupling case, high-quality CZ operation is obtained only in those regions where the conditional phase accumulation is sufficiently strong while the computational-state overlap remains large. However, since the relevant energy detunings vary substantially across the \((t_{0}^A,t_{0}^B)\) plane, a single fixed value of \(g\) cannot be optimal everywhere. If \(g\) is too large relative to the local detuning, excessive hybridization can reduce the overlap factor \(Q\) and increase leakage. Conversely, if \(g\) is too small, the conditional phase accumulation becomes slow, leading to longer CZ gate times. This motivates a locally adapted coupling strength, chosen here as \(g=\hbar|\omega^A - \omega^B|/200\), where the nominator is the local detuning and $\omega^{A,B}$ the individual qubit-splittings. This choice keeps the coupling in a perturbative regime while maintaining a finite conditional interaction across the parameter space. As a result, the tailored coupling improves the balance between state preservation and conditional phase accumulation, enlarging the region where high-\(Q\) CZ operation is possible, as shown in Fig. \ref{fig:tailoredCZ}.

\begin{figure}
    \centering
    \includegraphics{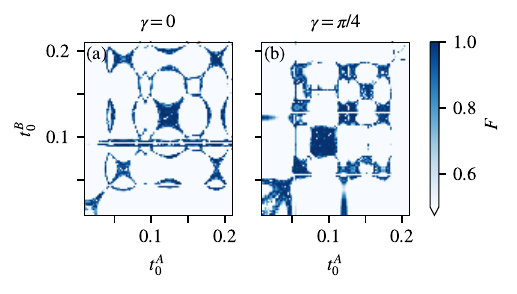}
    \caption{CNOT fidelity $F$ for (a) the $\gamma=0$ configuration and (b) the $\gamma=\pi/4$ configuration. The color minimum represents $F<0.5$. 
    }
    \label{fig:cnot}
\end{figure}

Finally, we also evaluate a CNOT gate driven via cross-resonance as a second example of two-qubit control. In contrast to the CZ gate, which is naturally associated with conditional phase accumulation under the capacitive interaction, a direct CNOT requires selective driving of conditional transitions and is therefore a more complex two-qubit gate in the sense that it is more sensitive to the detailed level structure and pulse parameters. Fig. \ref{fig:cnot} shows the CNOT fidelity in the \((t_{0}^A,t_{0}^B)\) plane for fixed coupling \(g/h=100~\mathrm{MHz}\) and fixed drive conditions. For each parameter point, the gate time is chosen from an analytical cross-resonance estimate of the conditional state-rotation time denoted $T^*$. This is an excellent approximation precisely in the high-fidelity regions, which are the ones of interest. The detailed procedure for the estimation of $T^*$ and treatment of ramp-up/down time is given in Appendix \ref{sec:cnotfidelity}. Our results show that the fidelity map is largely anticorrelated with the overlap factor \(Q\), as is reasonable for a cross-resonance-type mechanism. In regions with large \(Q\), the computational states are well preserved and only weakly hybridized. Consequently, the effective conditional drive is weak and the direct-CNOT fidelity is reduced. In contrast, stronger interaction-induced hybridization enhances the conditional response needed for CNOT operation, but also lowers \(Q\) by reducing the purity of the logical states and increasing leakage susceptibility. The direct-CNOT performance therefore reflects a compromise between computational-state preservation and conditional controllability, with the best operating points expected in an intermediate regime rather than at maximal \(Q\). Importantly, this should not be interpreted as a fundamental limitation of CNOT operation. The CNOT data shown here are not optimized over pulse shape, voltage-drive amplitude, pulse duration, or coupling strength \(g\). Optimizing these parameters, in particular by tailoring \(g\) to the local detuning as in the CZ analysis, is expected to enlarge the accessible high-fidelity region, although likely with a trade-off in gate time. A systematic optimization of direct CNOT operation is therefore left for future work.

\section{Concluding remarks}\label{sec:conclusion}

We have presented a comprehensive microscopic analysis of transmon qubits incorporating altermagnetic Josephson junctions, demonstrating that altermagnetism offers a compelling and largely untapped resource for qubit engineering. By computing the key performance metrics such as anharmonicity, decoherence rates, and gate operation times, we have shown that the altermagnetic band structure introduces a rich and tuneable parameter landscape.

Our central finding is that the transmon qubit, when hosted in an altermagnetic junction, exhibits a striking coexistence of strong decoherence protection and enhanced anharmonicity in the vicinity of $0-\pi$ transition points and within the $\phi$-state regime. This behavior originates from the dominance of higher harmonics in the energy-phase relation at these locations, which reshapes the potential profile of the qubit. We have further shown that the tradeoff between decoherence protection and fast gate times changes as the Néel field strength and the crystallographic orientation of the altermagnetic interfaces are varied. To tune the qubit in and out of its protected regime, we propose to exploit strain as an \textit{in-situ} tuning knob. Since mechanical strain modifies the effective altermagnetic order parameter and the relative crystal orientation at the junction interfaces \cite{khodas_prb_26}, it provides a reversible means to shuttle the qubit between its protected operating point, where coherence times are maximized, and a regime of faster gate operation where the anharmonicity landscape is less constraining. This dynamical tunability may be achieved using coupling to a gate-controlled piezoelectric substrate and represents an interesting alternative for qubit control compared to conventional transmon qubits. We have also considered entangled two-qubit (CZ) functionality, demonstrating that when the junction is close to distinct $0-\pi$ transitions or a $\phi$-state, high overlap factor and fast gate times emerge in the $d_{xy}$ crystallographic orientation of the altermagnet. Apart from the transmon, altermagnetic junctions offer the possibility of designing flux qubits and fluxonium operating at zero flux, due to their inherent $\pi$-state and large higher harmonics.

Several directions present themselves as natural extensions of the present work. On the materials side, the identification of concrete altermagnetic candidates with suitable superconducting proximity effect, controllable crystallographic texture at interfaces, and compatibility with established thin-film deposition processes, including strong coupling to a piezoelectric substrate, will be essential for experimental realization. On the theoretical side, it will be important to go beyond the ballistic transport regime employed here and to account for disorder at the altermagnetic-superconductor interface, which may partially lift the symmetry-imposed suppression of higher harmonics and thereby modify the predicted protection. The influence of altermagnetism on more exotic qubit modalities, such as gatemon architectures and hybrid semiconductor-superconductor qubits, also warrants further investigation \cite{vosoughinia_arxiv_25}.  More broadly, our results establish that the crystallographic degree of freedom inherent to altermagnets and the strength of the Néel order constitute a new design axis for superconducting quantum circuits. This suggests that the ongoing discovery and characterization of altermagnetic materials may have direct technological implications for fault-tolerant quantum computation, beyond their already recognized relevance to spintronics and topological physics. We hope this work motivates both experimental efforts to integrate altermagnetic materials into superconducting qubit platforms and further theoretical exploration of quantum coherence in unconventional magnetic systems.

\acknowledgments
J. Danon, T. Yamashita, K. Flensberg, A. Manesco and B. Lu are thanked for useful discussions. 
This work was supported by the Research Council of Norway
through Grant No. 353894 and its Centres of Excellence
funding scheme Grant No. 262633 “QuSpin.” Support from
Sigma2 - the National Infrastructure for High Performance
Computing and Data Storage in Norway, project NN9577K, is
acknowledged. \\

\textbf{Code and data availability:} The code and data used to prepare this manuscript are available
in Zenodo \cite{zenodo}.\\

\textbf{Disclosure of Generative AI use:} During the preparation of this manuscript, the authors used generative AI tools (Anthropic's Claude, Microsoft's Copilot, Google's Gemini and Z.ai's GLM 5.2) to generate, debug, and optimize numerical scripts used to plot the figures, as interactive brainstorming aids for discussing physical interpretations of the numerical results, and for style editing of portions of the text. All AI-generated outputs were thoroughly reviewed, verified, and edited by the authors, who maintain full responsibility for the scientific integrity, accuracy, and conclusions of this work.

\appendix

\section{Numerical procedure for computing energy levels with coupled qubits}
\label{sec:qubitoperation}

The numerical procedure  considers coupled qubits which individually have a Hamiltonian  of the form
\begin{align*}
H_i = 4E_C\hat{n}^2 + V(\varphi),
\end{align*}
where $\varphi$ is the phase and $\hat{n} = -i\partial_\varphi$ is the number operator. In principle, $V(\varphi)$ could be arbitrary, but here we have in mind superconducting qubits, with Josephson-like energies, for which $\varphi$ is interpreted as the phase difference across the junction. For concreteness, we here illustrate the numerical procedure for a standard transmon potential
\begin{align*}
V(\varphi) = -E_J\cos\varphi.
\end{align*}
but our methodology is valid for arbitrary $V(\varphi)$.
The above equation is a standard 1D problem in quantum mechanics, with periodic boundary conditions, $\varphi \in \left[0,2\pi\right]$. We discretize
\begin{align*}
\partial_\varphi^2 = \frac{\delta_{i,j+1} - 2\delta_{ij} + \delta_{i,j-1}}{\Delta\varphi^2},
\end{align*}
with $\Delta\varphi$ the spacing between "lattice points" for the phase-difference. Furthermore, we connect the lattice site at $\varphi_i = 0$ with $\varphi_{N-1} = 2\pi - \Delta\varphi$, allowing for hopping between these two lattice sites, in accordance with the periodic boundary conditions. Here, $N$ is the number of lattice points.\\

Upon diagonalization, we get $N$ energy levels with energies $\varepsilon_n$ and eigenvectors $\ket{n}$, for $n\in0,1,\ldots,N-1$. We truncate this system and retain only the $M$ lowest energy levels. For two (or more) decoupled qubits, we know immediately that the energy states are given as
\begin{align*}
\ket{nm} = \ket{n^{(a)}}\otimes\ket{m^{(b)}}\equiv\ket{l},
\end{align*}
for qubit $a$ and $b$, with energies
\begin{align*}
\varepsilon_{nm} = \varepsilon^{(a)}_n+\varepsilon^{(b)}_m \equiv \varepsilon_l.
\end{align*}
Hence, the decoupled Hamiltonian, projected onto these truncated energy states is a $2^M\times 2^M$ diagonal matrix,
\begin{align*}
H_{lk} = \langle l|H_1+H_2|k \rangle, 
\end{align*}
where we have assigned some ordering of the energy levels, indicated by index $l$. A reasonable convention is to order them in increasing order so that $l = 0$ is equal to $\varepsilon_{00}$, $l = 1$ is equal to $\varepsilon_{01}$, $l=2$ is equal to $\varepsilon_{10}$ and so on, giving the matrix structure 
\begin{widetext}
\begin{align*}\hat{H}_0=
\begin{pmatrix}
\varepsilon^{(a)}_0+\varepsilon^{(b)}_0 & 0   & 0   & \cdots & 0      & 0      \\
0   & \varepsilon^{(a)}_1+\varepsilon^{(b)}_0 & 0   & \cdots & 0      & 0      \\
0   & 0   & \varepsilon^{(a)}_0+\varepsilon^{(b)}_1 & \cdots & 0      & 0      \\
\vdots & \vdots & \vdots & \ddots & \vdots & \vdots \\
0   & 0   & 0   & \cdots & \varepsilon^{(a)}_{M}+\varepsilon^{(b)}_{M-1} & 0      \\
0   & 0   & 0   & \cdots & 0      & \varepsilon^{(a)}_M+\varepsilon^{(b)}_M
\end{pmatrix}.
\end{align*}
\end{widetext}
In effect, the ordering of the states in the eigenvector is 00, 01, 10, 11, and so on. The exact ordering depends on which is smaller. The point is that we can use one index to describe combinations of the two single-transmon energy levels.

\subsubsection{Coupling}
We assume that a capacitor couples the two qubits. This contributes the energy to the total system
\begin{align*}
E_g = \frac{1}{2}C_g\left(V_1 - V_2\right)^2 = \frac{1}{2}C_g\left(\dot{\Phi}_1 - \dot{\Phi}_2\right)^2,
\end{align*}
which gives a total Lagrangian
\begin{align*}
\mathcal{L} = \sum_i\frac{1}{2}C_i\dot{\Phi}_i^2 + \frac{1}{2}C_g\left(\dot{\Phi}_1 - \dot{\Phi}_2\right)^2 -\sum_iV_{i}(\Phi_i).
\end{align*}
The canonical momenta, $\partial\mathcal{L}/\partial\dot{\Phi}_i$, are now given as

\begin{align*}
\tilde{Q}_1 &= C_1\dot{\Phi}_1 + C_g\left(\dot{\Phi}_1 - \dot{\Phi}_2\right) = (C_1+C_g)\dot{\Phi}_1 - C_g\dot{\Phi}_2\\
\tilde{Q}_2 &= C_1\dot{\Phi}_2 - C_g\left(\dot{\Phi}_1 - \dot{\Phi}_2\right) = (C_2 + C_g)\dot{\Phi}_1 - C_g\dot{\Phi}_2
\end{align*}

We employ the basis $\Phi = \begin{pmatrix}\Phi_1 & \Phi_2 \end{pmatrix}^T$, so that
\begin{align*}
\tilde{Q} = M\dot{\Phi},
\end{align*}
with
\begin{align*}
M = \begin{pmatrix} C_1' & -C_g \\ -C_g & C_2' \end{pmatrix},
\end{align*}
where $C_i' = C_i + C_g$. Hence, $\dot{\Phi} = M^{-1}\tilde{Q}$. We calculate
\begin{align*}
M^{-1} = \frac{1}{C_1'C_2'-C_g^2}\begin{pmatrix} C_2' & C_g \\ C_g & C_1' \end{pmatrix} \equiv \begin{pmatrix} 1/C_1^{''} & 1/C_g^{''} \\ 1/C_g^{''} & 1/C_2^{''} \end{pmatrix},
\end{align*}
with

\begin{align*}
C_1^{''} &= \left(C_1C_2 + C_g(C_1+C_2)\right)/(C_2+C_g) = C_1 + \frac{C_gC_2}{C_g+C_2} \\
C_2^{''} &= \left(C_1C_2 + C_g(C_1+C_2)\right)/(C_1+C_g) = C_2 + \frac{CgC_1}{C_g+C_1} \\
C_g^{''} &= \left(C_1C_2 + C_g(C_1+C_2)\right)/C_g = C_1+C_2 + \frac{C_1C_2}{C_g}
\end{align*}

We can in principle couple as many qubits as we like in this way. The kinetic energy is thus expressible as 
\begin{align*}
T = \frac{1}{2}\tilde{Q}M^{-1}\tilde{Q} = \frac{\tilde{Q}^2_1}{2C_1^{''}} + \frac{\tilde{Q}_2^2}{2C_2^{''}} + \frac{\tilde{Q}_1\tilde{Q_2}}{C_g^{''}}.
\end{align*}
This can be seen as follows. The first two terms in $\mathcal{L}$ is $\frac{1}{2}\dot{\Phi}^T M \dot{\Phi}$, which defines $M$ as given above. This is verified by a direct calculation, now with $\Phi$ as a vector with $\Phi_i$ as elements. Furthermore, the canonical momenta also satisfy $\tilde{Q} = M\dot{\Phi}$, as may also be verified by direct calculation. Here $\tilde{Q}$ is a vector with elements $\tilde{Q}_i$. Expressing $\frac{1}{2}\dot{\Phi}^T M \Phi$ in terms of $\tilde{Q}$ is what gives $T$, identified as the kinetic energy because it contains the canonical momenta squared. The Hamiltonian is thus
\begin{align*}
H = \frac{\tilde{Q}^2_1}{2C_1^{''}} + \frac{\tilde{Q}_2^2}{2C_2^{''}} + \frac{\tilde{Q}_1\tilde{Q_2}}{C_g^{''}} + V_{1}(\Phi_1) + V_{2}(\Phi_2),
\end{align*}
or in terms of number operator and phase,
\begin{align*}
H = H_1 + H_2 + g n_1n_2,
\end{align*}
with $g = 8\sqrt{E_{C,1}E_{C,2}C_1''C_2''}/C_g''$. We get the coupling term, 
\begin{align*}
V_c =-g\partial_{\varphi_1}\partial_{\varphi_2} =  -g\partial_{\varphi}\otimes\partial_\varphi,
\end{align*}
now expressed in terms of the phase difference $\varphi_i$ of Josephson junction $i$, which is proportional to the flux through the junction, $\Phi_i$. We express $\varphi_1$ and $\varphi_2$ in terms of the same coordinate $\varphi$, and a Kronecker product. This is similar in spirit as the equivalence of writing an integral over $x$ squared as a double integral over $x$ and $y$. We underline that the phase differences for the two qubits are in general not the same.
Next, we project this potential onto the subspace of the truncated energy levels,
\begin{align*}
V_{lk} = \langle l|V_c|k \rangle,
\end{align*}
with matrix structure $\hat{V}$. At this point we have a coupled system of two qubits, each with $M$ levels whose Hamiltonian is expressed in the decoupled basis,
\begin{align*}
\hat{H} = \hat{H}_0 + \hat{V}.
\end{align*}
If we diagonalize this, to obtain the diagonal matrix $\tilde{H}$, we get the system expressed in the dressed basis. Indeed, the eigenvectors $v_n$ are $2^M\times 1$ arrays, and give the dressed eigenstate as a linear combination of decoupled basis vectors,
\begin{align*}
\ket{l}_d = \sum_{j=0}^M v_l(j)\ket{j}.
\end{align*}
The overlap matrix emerges immediately by placing the elementwise absolute value of the eigenvectors $v_n$ as columns,
\begin{align*}
O = \begin{pmatrix} |v_1|^2 & |v_2|^2 & \ldots & |v_{2^M}|^2 \end{pmatrix}. 
\end{align*}
Note that we here mean elementwise squared absolute values, not scalar products with itself. In effect,  $\nu_1$ is a $2^M\times 1$ column vector
\begin{align}
    \nu_1 = \begin{pmatrix}
        \nu_1(j=1)\\
        \nu_1(j=2)\\
        \ldots \\
        \nu_1(j=2^M) \\
    \end{pmatrix}
\end{align}
with possibly complex entries $\nu_1(j)$. This distinction is important because the scalar product $v_n^\dagger v_n = 1$ by the normalization of the eigenvectors. To be concrete, consider for instance $M = 2$. The two lowest energy levels of each qubit are then included. Each energy level has an eigenvector which is an array of $N$ elements (the number of points in the $\varphi$ lattice). The decoupled basis vectors $\ket{l} = \ket{n}^{(a)}\otimes\ket{m}^{(b)}$ have $N^2$ elements. The reduced Hamiltonian $\hat{H}$, on the other hand is $4 \times 4$. Diagonalizing this, one obtains eigenvectors with 4 elements, $\ket{l}_d$. They express the weight that each of the basis vectors $\ket{l}$ contribute with. If, say the second to lowest dressed eigenvalue has eigenvector $v_2 = \begin{pmatrix}0 & 1/\sqrt{2} & 1/\sqrt{2} & 0\end{pmatrix}$, this means that the given dressed state is an equal superposition of the second and third decoupled eigenvectors. That is exactly the definition of the overlap matrix.  $O$ is generally a $2^M \times 2^M$ matrix.

\subsubsection{Driving}
We can introduce qubit driving, allowing for gate operations, by the potential 
\begin{align*}
V_d(t) = V_0^As_a(t)\sin(\Omega_at)\; n_a + V_0^Bs_b(t)\sin(\Omega_bt) n_b,
\end{align*}
with $V_0^i$ the magnitude of the driving, as applied to qubit $i$, $s_i(t)$ is a time-dependent ramping potential and $\Omega_i$ is the driving frequency. This is equal to 
\begin{align}
V_d(t) &= -i\Big[ V_0^As_a(t)\sin\Omega_a t\; \partial_\varphi\;\otimes \hat{I} \notag\\
&+ V_0^Bs_b(t)\sin\Omega_b t\;\hat{I}\otimes \partial_\varphi \Big],
\end{align}
where we have used that $n_a = -i\partial_{\varphi_a} = -i\partial_\varphi\otimes \hat{I}$, and $n_b = -i\partial_{\varphi_b} = \hat{I}\otimes\partial_\varphi$, now referring to the two qubits as $a$ and $b$ rather than 1 and 2. Once again, everything is expressed in terms of the variable $\varphi$, which is discretized to a 1D lattice of $N$ points. The derivative operator then becomes an $N\times N$ tridiagonal matrix, which also determines the size of the Hilbert space of our individual qubits, and therefore also the size of the identity matrix. Whether an operator acts on qubit $a$ or $b$ determines whether it is on the left or the right of the Kronecker product $\otimes$.  We project this operator onto the decoupled basis,
\begin{align*}
V_{d,nm}(t) = \langle n|V_d(t)|m\rangle,
\end{align*}
which reduces its size to $2^M\times 2^M$, with matrix structure $\hat{V_d}(t)$. Next, we project this Hamiltonian onto the dressed basis, 
\begin{align*}
\tilde{V}_{d,lk}(t) = v_l^\dagger \hat{V}_d(t) v_k.
\end{align*}
The driven Hamiltonian in the dressed basis is thus
\begin{align*}
\tilde{H}_{tot}(t) = \tilde{H} + \tilde{V}_d(t). 
\end{align*}
To find the time evolution, we employ the Suzuki-Trotter decomposition. We pick an initial state $v(t=0)$, that is, a superposition of the dressed eigenvectors as determined by the way we initialize our system, as our starting point. The time evolution operator is given as 
\begin{align}
U(t_0,t_n) &= \mathcal{T}\exp\left[-i\int_{t_0}^{t_n} dt\;H(t)\right] \notag\\
&= \mathcal{T}\exp\left[-i\tilde{H}(t_n-t_0) - i\int_{t_0}^{t_n} dt\;\tilde{V}_d(t)\right].
\end{align}
We discretize time in steps of $\Delta t$, which turns the integral into a sum. Furthermore, for sufficiently small time-steps, we may assume that all terms in the exponent commute. This, in turn, allows us to write the sum in the exponent as a product over exponentials at each time step. In this way, we  calculate the time-evolution operator as 
\begin{align*}
U(0,t = K\Delta t) = \prod_{k=0}^{K} e^{-i\tilde{H}\Delta t/2}\;e^{-i\tilde{V}_d(k\Delta t)\Delta t} e^{-i\tilde{H}\Delta t/2},
\end{align*}
where $\tilde{H}$ is placed symmetrically for some added numerical accuracy. This can be seen from the Baker-Campbell-Hausdorff expansion,
\begin{align*}
e^{X}e^{Y} = e^{Z},
\end{align*}
with
\begin{align*}
Z = X + Y + \frac{1}{2}\left[X,Y\right] + \frac{1}{12}\left[X,\left[X,Y\right]\right]+\ldots
\end{align*}
Consider now the product
\begin{align*}
M = e^{-i\tilde{H}\Delta t/2}\;e^{-i\tilde{V}_d(k\Delta t)\Delta t} e^{-i\tilde{H}\Delta t/2},
\end{align*}
and apply this formula to the last two terms,
\begin{align}
M &= e^{-i\tilde{H}\Delta t/2} \exp\Big[-i\tilde{V}_d(k\Delta t)\Delta t - i\tilde{H\Delta t/2} \notag\\
&- \frac{1}{2}\left[\tilde{V}_d(k\Delta t)\;,\;\frac{1}{2}\tilde{H}\right]\Delta t^2 + O(\Delta t^3)\Big].
\end{align}
Now apply this formula to the remaining two exponentials,
\begin{align}
M &= \exp\Big(-i\tilde{V}_d(k\Delta t)\Delta t - i\tilde{H\Delta t} -\frac{1}{2}\left[\frac{1}{2}\tilde{H}\;,\;\tilde{V}_d(k\Delta t)\right]\Delta t^2 \notag\\
&- \frac{1}{2}\left[\tilde{V}_d(k\Delta t)\;,\;\frac{1}{2}\tilde{H}\right]\Delta t^2 + O(\Delta t^3)\Big) 
\\&= \exp\Big(-i\tilde{V}_d(k\Delta t)\Delta t - i\tilde{H}\Delta t + O(\Delta t^3)\Big).
\end{align}
The error in the exponent is of order $O(\Delta t^3)$ when doing the Suzuki-Trotter decomposition symmetrically, as opposed to being of order $O(\Delta t^2)$ when asymmetric. The change in our initial state after time $t = K\Delta t$ is finally given as
\begin{align*}
v(t) = U(0,t)v(0).
\end{align*}

We have verified that our numerical results agree when diagonalizing the qubit Hamiltonian both when using a phase basis, as described in detail above, or in a charge basis. For dephasing in particular, however, using a charge basis gives much more accurate results even when truncating the number of basis states at smaller value than in the phase basis case.

\section{CNOT fidelity calculation}\label{sec:cnotfidelity}
Consider a Hamiltonian of two decoupled qubits, with a driving potential applied to qubit A,
\begin{align}
H = \begin{pmatrix} E_{00} & V(t) & 0 & 0 & \\
V^*(t) & E_{10} & 0 & 0 \\
0 & 0 & E_{01} & V(t) \\
0 & 0 &V^*(t) & E_{11}
\end{pmatrix}.
\end{align}
Here, $V(t) = \braket{00|\hat{V}(t)|10} = \braket{01|\hat{V}(t)|11}$. Because the qubits are decoupled, there are only transitions between $\ket{0}_A\to\ket{1}_A$ and vice versa, qubit B is unchanged. If a capacitive coupling exists between the qubits, the states hybridize. To capture the essential physics, we approximate the capacitive interaction
between the two qubits by first projecting
\[
H_{\mathrm{int}} = g \hat n_A \hat n_B
\]
onto the computational subspace. Retaining the dominant transverse matrix
elements (the diagonal matrix elements are typically suppressed by parity symmetry) of the charge operators gives an effective interaction
\[
H_{\mathrm{int}} \simeq J \sigma_x^A \sigma_x^B .
\]
Expanding \(\sigma_x=\sigma_+ + \sigma_-\), this interaction contains both
excitation-conserving terms,
\[
\sigma_+^A\sigma_-^B + \sigma_-^A\sigma_+^B,
\]
which couple \(|10\rangle\) and \(|01\rangle\), and counter-rotating terms,
\[
\sigma_+^A\sigma_+^B + \sigma_-^A\sigma_-^B,
\]
which couple \(|00\rangle\) and \(|11\rangle\). Within the rotating-wave approximation applied to the static interqubit
coupling, we neglect the counter-rotating terms
\(\sigma_+^A\sigma_+^B\) and \(\sigma_-^A\sigma_-^B\). These terms couple \(|00\rangle\) and \(|11\rangle\), whose energy separation is
approximately \(\hbar(\omega_A+\omega_B)\). In the interaction picture, the
corresponding coupling terms therefore acquire phase factors oscillating at
\(\omega_A+\omega_B\). 
For example, the first-order perturbation theory transition amplitude generated by such a term is
bounded by
\begin{align}
A_{00\rightarrow 11}(t)
&\sim
-\frac{iJ}{\hbar}\int_0^t dt'\, e^{i(\omega_A+\omega_B)t'}
\notag\\
&=
\frac{J}{\hbar(\omega_A+\omega_B)}
\left[1-e^{i(\omega_A+\omega_B)t}\right],
\end{align}
and is therefore of order
\[
|A_{00\rightarrow 11}|
\lesssim
\frac{J}{\hbar(\omega_A+\omega_B)}.
\]
Thus, when \(\omega_A+\omega_B\gg J/\hbar\), the counter-rotating terms only
produce a small virtual admixture and can be neglected within the RWA. By contrast, the excitation-conserving
terms \(\sigma_+^A\sigma_-^B\) and \(\sigma_-^A\sigma_+^B\) couple
\(|10\rangle\) and \(|01\rangle\), whose energy mismatch is only
\(\hbar(\omega_A-\omega_B)\), and are therefore retained. The remaining exchange interaction conserves
excitation number and only hybridizes the single-excitation subspace. Thus, to leading order within this first RWA, the coupled eigenstates may be written as
\[
|0\rangle = |00\rangle ,
\]
\[
|1\rangle =
\cos\theta |10\rangle
+
\sin\theta |01\rangle ,
\]
\[
|2\rangle =
-\sin\theta |10\rangle
+
\cos\theta |01\rangle ,
\]
\[
|3\rangle = |11\rangle .
\]
The mixing angle \(\theta\) depends on the strength of the coupling $g$. This approximation provides the dressed
basis used below to estimate the resonant microwave-driven
\(|1\rangle\leftrightarrow |3\rangle\) Rabi frequency and hence the idealized
CNOT gate time. Now, the Hamiltonian in the hybridized eigenbasis $\{\ket{0},\ket{1},\ket{2},\ket{3}\}$ becomes
\begin{align}
H = \begin{pmatrix} E_{0} & V(t)\cos\theta & -V(t)\sin\theta & 0 & \\
V^*(t)\cos\theta & E_{1} & 0 & V(t)\sin\theta \\
-V^*(t)\sin\theta & 0 & E_{2} & V(t)\cos\theta \\
0 & V^*(t)\sin\theta &V^*(t)\cos\theta & E_{3}
\end{pmatrix}.
\end{align}
We assume the following form of the potential,
\begin{align}
V(t) = S(t)V_0\cos\omega_d t,
\end{align}
where $V_0$ is its amplitude, $S(t)$ is an envelope function, and $\omega_d$ is the drive frequency. We write the Hamiltonian as $H_0 + H_1(t)$. In the interaction picture, only the latter part contributes to the time evolution, and we get
\begin{widetext}
\begin{align}
H_{1,I}(t)
&=
e^{iH_0t/\hbar} H_1(t) e^{-iH_0 t/\hbar}
\nonumber \\
&=
\begin{pmatrix}
0
& V(t)e^{-i\omega_{01}t}\cos\theta
& -V(t)e^{-i\omega_{02}t}\sin\theta
& 0
\\
V^*(t)e^{i\omega_{01}t}\cos\theta
& 0
& 0
& V(t)e^{-i\omega_{13}t}\sin\theta
\\
-V^*(t)e^{i\omega_{02}t}\sin\theta
& 0
& 0
& V(t)e^{-i\omega_{23}t}\cos\theta
\\
0
& V^*(t)e^{i\omega_{13}t}\sin\theta
& V^*(t)e^{i\omega_{23}t}\cos\theta
& 0
\end{pmatrix}.
\end{align}
Here we define positive transition frequencies
\[
\omega_{ij} = \frac{E_j-E_i}{\hbar},
\qquad i<j,
\]
for the independent transitions $(i,j)=(0,1),(0,2),(1,3),(2,3)$ where \((H_{1,I})_{ij}=\langle i|H_{1,I}|j\rangle\). We assume the usual weak-coupling regime
in which the dressed states can be ordered as \(E_0<E_1,E_2<E_3\). Next we employ a second RWA associated with the microwave drive. We write the
potential as
\begin{align}
V(t)
=
\frac{1}{2} S(t) V_0
\left(e^{i\omega_d t} + e^{-i\omega_d t}\right).
\end{align}
Each interaction-picture matrix element then contains one term oscillating at
the detuning
\[
\Delta_{ij}=\omega_{ij}-\omega_d,
\]
and one counter-rotating term oscillating at \(\omega_{ij}+\omega_d\). Since
we choose the drive frequency \(\omega_d\) close to a particular transition,
the terms oscillating at \(\omega_{ij}+\omega_d\) are rapidly rotating compared
with the near-resonant terms and are discarded. This gives
\begin{align}
\tilde{H}_{1,I}(t)
=
\frac{1}{2}V_0S(t)
\begin{pmatrix}
0
& e^{-i\Delta_{01}t}\cos\theta
& -e^{-i\Delta_{02}t}\sin\theta
& 0
\\
e^{i\Delta_{01}t}\cos\theta
& 0
& 0
& e^{-i\Delta_{13}t}\sin\theta
\\
-e^{i\Delta_{02}t}\sin\theta
& 0
& 0
& e^{-i\Delta_{23}t}\cos\theta
\\
0
& e^{i\Delta_{13}t}\sin\theta
& e^{i\Delta_{23}t}\cos\theta
& 0
\end{pmatrix}.
\end{align}
We drive this system precisely at the resonance
\[
\omega_d=\omega_{13},
\]
so that \(\Delta_{13}=0\). The interaction-picture Hamiltonian then becomes 
\begin{align}
\tilde{H}_{1,I}(t)
=
\frac{1}{2}V_0S(t)
\begin{pmatrix}
0
& e^{-i\Delta_{01}t}\cos\theta
& -e^{-i\Delta_{02}t}\sin\theta
& 0
\\
e^{i\Delta_{01}t}\cos\theta
& 0
& 0
& \sin\theta
\\
-e^{i\Delta_{02}t}\sin\theta
& 0
& 0
& e^{-i\Delta_{23}t}\cos\theta
\\
0
& \sin\theta
& e^{i\Delta_{23}t}\cos\theta
& 0
\end{pmatrix}.
\end{align}
\end{widetext}
A well-functioning CNOT gate is obtained if all other transitions except $\ket{1}\to\ket{3}$ are far off resonance when driven at this frequency, in which case the phase factors present in all other terms are suppressed by the same argument as the RWA. If each individual qubit were to contain only two energy levels, as in this treatment, $\Delta_{02}$ would equal $\Delta_{13}$ and a CNOT gate would be impossible. However, in the actual numerics where higher-lying energy levels are retained, virtual transitions beyond the qubit subspace will in general ensure that $\Delta_{02}\neq\Delta_{13}$. In the resonant regime for CNOT-functionality, one therefore has the approximate Hamiltonian
\begin{align}
\tilde{H}_{1,I}(t) = \frac{1}{2}V_0S(t)
\begin{pmatrix} 0 & 0 & 0 & 0 & \\
0 & 0 & 0 & \sin\theta \\
0 & 0 & 0 & 0 \\
0 & \sin\theta &0 & 0
\end{pmatrix}.
\end{align}
This reduces to a two-level effective Hamiltonian in the two-level subspace spanned by $\{\ket{1},\ket{3}\}$:
\begin{align}
H_\text{eff}(t) = \frac{1}{2}V_0S(t)\sin\theta\;\sigma_x.
\end{align}
The time evolution of this Hamiltonian is simply 
\begin{align}
U(t) = \exp\left[-\frac{i}{\hbar}\int_0^t dt'\; H_\text{eff}(t')\right] = e^{-i\Theta(t)\sigma_x/2}.
\end{align}
This is a spin rotation matrix, with angle
\begin{align}
\Theta(t) = \frac{V_0}{\hbar}\sin\theta\int_0^t dt'\; S(t').
\end{align}
 A transition has taken place once an angle of $\pi$ has been reached. We refer to this time as $T_R$. For a constant envelope function, $S(t) = 1$, we get
\begin{align}
T_R = \frac{\pi\hbar}{V_0\sin\theta}.
\end{align}
In the full numerical procedure $V\sin\theta$ is replaced by the relevant matrix element
\[
V_{13} \equiv \langle 1|\hat V_0|3\rangle ,
\]
where the drive is written as
\[
\hat V(t)=S(t)\hat V_0\cos(\omega_d t).
\]. To avoid ringing and other artifacts of this instantaneous switching on and off, it is generally better to use an envelope function that ramps up at the beginning and down at the end of some period $T^*$. A common choice is the sine-squared envelope,
\begin{align}
S(t) = \begin{cases} \sin^2\frac{\pi t}{2aT^*},&  0 <t < aT^* \\
1, & aT^* < t < (1-a)T^* \\
\sin^2\frac{\pi(T^*-t) }{2aT^*}, &(1-a)T^* < t < T^*
\end{cases}
\end{align}
where $0<a<\frac{1}{2}$. Since the rotation angle \(\Theta(t)\) is proportional to the pulse area
\(\int_0^t dt'\,S(t')\), the shaped pulse enacts the same rotation as the
square pulse when its area equals the square-pulse duration \(T_R\). Therefore,
\begin{align*}
T_R &= \int_0^{T^*}dt'\; S(t') \\
&= \int_0^{aT^*}dt'\; \sin^2\frac{\pi t'}{2aT^*} + \int_{aT^*}^{(1-a)T^*}dt'\; \\
&+ \int_{(1-a)T^*}^{T*}dt'\sin^2\frac{\pi (T^*-t')}{2aT^*} = (1-a)T^*,
\end{align*}
from which we identify
\begin{align}
T^* = \frac{T_R}{1-a}. 
\end{align}
In other words, with a non-constant envelope, the system must be driven for slightly longer than for a constant envelope.

In the main manuscript, the ramp-up period is defined by $a=1/10$ and the fidelity is thus evaluated at $T^* \simeq 1.1 T_R$. This is a valid procedure in the high-fidelity regions, since that is precisely where the CNOT-dynamics is expected to work as described in this appendix. To validate this procedure, we also performed a direct search over a gate time in the interval $0 < t < T^*$. This did not reveal additional high-fidelity regions beyond those identified using the analytical $T_R$ estimate. This shows that the main structure of the fidelity map provided in the main manuscript is not an artifact of the chosen gate time prescription. 

\section{Analytical expression for the free energy of an altermagnetic Josephson junction}
\label{sec:analytical}

We assume momentum parallel to the interface with the superconductors, $k_y$, is conserved throughout the system and write the altermagnetic Hamiltonian as
\begin{align}
H = \frac{\hbar^2}{2m}k^2 + \frac{\hbar^2\beta}{2m}\left[2\cos\gamma k_xk_y - \sin\gamma (k_x^2-k_y^2) \right] - \mu
\end{align}
where $\beta$ is the dimensionless strength of the altermagnetism, and $\gamma$ is the orientation of the band structure lobes in the $xy$ plane. We here use $\beta=2t_0$ to denote the altermagnetic strength to avoid confusion with the tunneling amplitude that will be used later on in this section. $\gamma = 0$ has lobes along the $k_x$ and $k_y$ axes, and $\gamma = \pi/4$ has them oriented diagonally. Solving the eigenvalue problem $H\psi = E\psi$ in the limit $E/\mu \ll 1$, gives the wavevector $k_{x,s}$, for spin $s$,
\begin{align*}
k_{x,s} = \frac{-s\beta k_y\cos\gamma+\eta\sqrt{k_\fermi^2\left(1-s\beta\sin\gamma\right)-(1-\beta^2)k_y^2}}{1-s\beta\sin\gamma},
\end{align*}
with $\eta\pm1$. We choose $\eta = +1$ for electrons moving from left to right. In the limit of weak altermagnetism, $\beta\ll 1$, $k_{x,s}$ simplifies to
\begin{align}
k_{x,s} = k_\fermi\left[-s\beta\cos\gamma\sin\theta +\left(\cos\theta - s\beta\sin\gamma\frac{\cos 2\theta}{2\cos\theta}\right)\right],
\end{align}
where we have defined the angle of incidence $\sin\theta = k_y/k_\fermi$.
A particle that travels at such an angle $\theta$ through the altermagnetic barrier, whose thickness along the $x$ axis is $d$, can therefore be assumed to accumulate a phase $k_{x,s} d$. In Josephson junctions, such a phase is important and we can take it into account in a tunneling Hamiltonian approach by including it in the tunneling amplitude,
\begin{align}
t_s(\theta) = t_0 e^{ik_{x,s}d}
\label{eq:ts}
\end{align}
We proceed with the tunneling Hamiltonian approach, which for general spin-dependent tunneling takes the form
\begin{align}
H_T = \sum_{kqs}\left[t_{kqs} c_{ks,R}^\dagger c_{qs,L} + t_{kqs}^*c_{qs,L}^\dagger c_{ks,R}\right].
\end{align}
The superconductors are taken as conventional Bardeen-Cooper-Schrieffer (BCS) ones, to which $H_T$ constitutes a perturbation. In the imaginary time interaction picture, we have
\begin{align}
H_0 &= H_L + H_R \\
H_T &= e^{\tau H_0} H_T e^{-\tau H_0}.
\end{align}
In Nambu space, the tunneling Hamiltonian becomes
\begin{align}
H_T = \sum_{kq}\left[\psi_{k,R}^\dagger\hat{T}_{kq}\psi_{q,L} + \psi_{q,L}^\dagger\hat{T}_{kq}^\dagger\psi_{q,R}\right],
\end{align}
with $\psi_{k,X} = \begin{pmatrix} c_{k\uparrow,X} & c_{k\downarrow,X} & c^\dagger_{-k\uparrow,X} & c^\dagger_{-k\downarrow,X}\end{pmatrix}^T$, and 
\begin{align}
\hat{T}_{kq} = \begin{pmatrix} t_{kq\uparrow} & 0 & 0 & 0 \\
	0 & t_{kq\downarrow} & 0 & 0 \\
	0 & 0 & -t_{\bar{k}\bar{q}\uparrow}^* & 0 \\
	0 & 0 & 0 & -t_{\bar{k}\bar{q}\downarrow}^*
\end{pmatrix},
\end{align}
with $\bar{k} = -k$. From this we may find the partition function as
\begin{align}
\frac{Z}{Z_0} = \left\langle \;T_\tau\exp\left(-\int_0^\beta d\tau\; H_T(\tau)\right)\right\rangle_0.
\label{eq:part}
\end{align}
The subscript $0$ means that we should calculate the expectation value in the unperturbed system. From this we find the free energy,
\begin{align}
F = F_0 - \frac{1}{\beta}\ln\left(\frac{Z}{Z_0}\right).
\label{eq:fedef}
\end{align}
We ignore $F_0$ as the background energy does not contribute to the supercurrent so long as it is phase-independent. Furthermore, since everything is diagonal in spin, we can treat each spin species separately, so that
\begin{align}
F = \sum_s F_s = F_\uparrow + F_\downarrow.
\end{align}
We proceed with finding $F_s$. In general, the free energy is calculated from the series expansion of \cref{eq:fedef}, which is equivalent to the series expansion of \cref{eq:part}, while retaining only the connected diagrams, per the linked cluster theorem, 
\begin{align}
F = -\frac{1}{\beta}\sum_{n=0}^\infty\frac{(-1)^n}{n!}\prod_{j=1}^n\int_0^\beta d\tau_j\;\left\langle T_\tau H_T(\tau_1)\ldots H_T(\tau_n)\right\rangle_0.
\end{align}
By meticulous application of Wick's theorem, it can be shown that the contribution of order $2n$ can be written as
\begin{align}
F_s^{(2n)} = \frac{1}{n\beta}\sum_{\omega_n}\Re\text{Tr}\left[\left(\sum_{kq}\hat{T}_{k_1q_1}\hat{G}_L(q,i\omega_n)\hat{T}^\dagger_{k_1q_1}\hat{G}_{R},(k,i\omega_n)\right)^n\right],
\end{align}
where we have suppressed the index $s$ on the right hand side for brevity, and
\begin{align}
\hat{G}_{Rda}(k,\tau_1-\tau_2) &= -\left\langle T_\tau \psi_{k,Rd}(\tau_2)\psi^\dagger_{k,Ra}(\tau_1)\right\rangle_0, \\
\hat{G}_{Lbc}(k,\tau_2-\tau_1) &= -\left\langle T_\tau \psi_{k,Lb}(\tau_1)\psi^\dagger_{k,Lc}(\tau_2)\right\rangle_0,
\end{align}
are Green functions for the right and left side, respectively, written here with Nambu indices $a$, $b$, $c$, and $d$. Next, we assume that the tunneling amplitude is flux-conserving, and has the form
\begin{align}
t_{kqs} = t_s(\theta)\sqrt{v_{\fermi x,L}(\theta) v_{\fermi x, R}(\theta)} = t^+_s(\theta) v_{\fermi x}(\theta),
\end{align}
where $v_{\fermi x}(\theta)$ is the Fermi velocity normal to the interface incidence angle $\theta$, assumed identical in the two superconductors. Converting the sums over momentum to integrals, with $\xi_k = \hbar^2(k_x^2+k_y^2-k_F^2)/2m \simeq \hbar v_{Fx}(k_y)(k_x-k_F)$, and $v_{Fx}(k_y) = \hbar\sqrt{k_F^2-k_y^2}$ gives
\begin{align*}
\sum_{k_x} &= \frac{d}{2\pi}\int d\xi_k\; \frac{m}{\hbar}\left[\frac{2m}{\hbar}\xi_k + k_F^2-k_y^2\right]^{-1/2}\\
&\simeq \frac{d}{2\pi}\int d\xi_k\; \frac{m}{\hbar\sqrt{k_F^2-k_y^2}}\\ &=\frac{d}{2\pi\hbar v_{Fx}(k_y)}\int d\xi_k.
\end{align*}
Hence, if the two superconductors are of identical size, we get
\begin{align}
F_s^{(2n)} =\frac{1}{n\beta}\frac{d^2}{4\pi^2\hbar^2}\sum_{\omega_n}\sum_{k_y}\;\Re\text{Tr}\left[\left(M\right)^n\right],
\end{align}
with  
\begin{align*}
M = \int d\xi_k\int d\xi_q\;\hat{\Gamma}(\theta)\hat{G}_L(\xi_q,i\omega_n)\hat{\Gamma}^\dagger(\theta)\hat{G}_{R}(\xi_k,i\omega_n),
\end{align*}
and 
\begin{align*}
\hat{\Gamma}(\theta) = \begin{pmatrix} t_{s}(\theta) & 0  \\
	 0 & -t^*_{\bar{s}}(\theta+\pi)
\end{pmatrix}.
\end{align*}
We use a gauge where the gap is real, in which case the Green function for spin $s$ becomes
\begin{align}
\hat{G}_X(\xi_k,i\omega_n) = \frac{1}{\omega_n^2+\xi_k^2+\Delta^2}\begin{pmatrix}i\omega_n+\xi_k & s\Delta \\ s\Delta & i\omega_n - \xi_k \end{pmatrix}.
\end{align}
Integration over $\xi_k$ then gives
\begin{align}
\int d\xi_k\;\hat{G}_X = \frac{\pi}{\sqrt{\omega_n^2+\Delta^2}}\begin{pmatrix}i\omega_n & s\Delta \\ s\Delta & i\omega_n \end{pmatrix}\equiv \hat{g}_X,
\end{align}
reducing $M$ to
\begin{align}
M = \hat{\Gamma}(\theta)\hat{g}_L(i\omega_n)\hat{\Gamma}^\dagger(\pi-\theta)\hat{g}_{R}(i\omega_n).
\end{align}
Next, we sum over $n$, using the identity $\text{Tr}\ln\left(1-A\right) = -\sum_{n=1}^\infty \text{Tr}(A^n)/n$, giving
\begin{align}
F_s &= \sum_{n=1}^{\infty} F_s^{(2n)} \nonumber\\
&= - \frac{1}{\beta}\sum_{k_y}\sum_{\omega_n}\Re\text{Tr}\ln\left[1 - \left(\frac{d}{2\pi\hbar}\right)^{2}M\right]. 
\end{align}
Furthermore, $\text{Tr}\ln(A) = \ln\det(A)$, so that
\begin{align}
F_s = = - \frac{1}{\beta}\sum_{k_y}\sum_{\omega_n}\Re\ln\det\left[1 - \left(\frac{d}{2\pi\hbar}\right)^{2}M\right]. 
\end{align}
We can calculate the determinant, which gives

\begin{widetext}
\begin{align}
F_s = -\frac{1}{\beta}\sum_{k_y}\sum_{\omega_n}\Re\ln \left[\frac{\omega^2\left(1+|\tilde{t}(k_y)|^2\right)^2 + \Delta^2\left[(1+|\tilde{t}(k_y)|^2)^2 - 2\left(|\tilde{t}(k_y)|^2 - \Re(\tilde{t}(k_y)\tilde{t}(-k_y))\right)\right]}{\omega^2+\Delta^2}\right],
\end{align}
where we have defined $\tilde{t} = d t/2\hbar$. 
\end{widetext}
We discard the denominator as it is independent of the phase. We also divide everything by $(1+|\tilde{t}|^2)^2$, another phase-independent contribution. Next, we assume zero temperature, in which case the Matsubara sum becomes an integral, 
\begin{align*}
\frac{1}{\beta}\sum_{\omega_n}\to \int\frac{d\omega}{2\pi},
\end{align*}
so that we get
\begin{align}
F_s = -\sum_{k_y} \int\frac{d\omega}{2\pi}\Re\ln\left(\omega^2 + a^2\right),
\end{align}
with
\begin{align*}
a^2 = \Delta^2\left[1 - \frac{2\left(|\tilde{t}(k_y)|^2 - \Re(\tilde{t}(k_y)\tilde{t}(-k_y))\right)}{(1+|\tilde{t}(k_y)|^2)^2}\right].
\end{align*}
To solve this integral, we use cut-off regularization,
\begin{align*}
I &= \int_{-\zeta}^{\zeta}\frac{d\omega}{2\pi}\;\ln(\omega^2+a^2) \\
&= \frac{1}{2\pi}\left[4a\arctan\frac{\zeta}{a} - 4\zeta + 2\zeta\ln(a^2+\zeta^2)\right].
\end{align*}
Letting $\zeta\to \infty$. Then $\arctan(\zeta/a)\to\pi/2$, and $\ln(a^2+\zeta^2)\to 2\ln\zeta$, the integral approaches
\begin{align*}
I\to a - \frac{2\zeta}{\pi}+\frac{2\zeta\ln\zeta}{\pi}.
\end{align*}
Hence, there is a divergent background to our free energy, but also a cut-off independent part that depends on the superconducting phase difference from which we identify the regularized integral as $I_{R} = a$. Additionally, since the $\zeta$-terms are independent of the
superconducting phase, they do not affect either the Josephson
current or the qubit Hamiltonian in any case, since they only add a constant
proportional to the identity operator.
\\

Finally, we use \cref{eq:ts}, assuming that all $k_y$-dependence occurs only in the phase of the tunneling amplitude, and thus find
\begin{align}
F_s = -\sum_{k_y}\Delta\sqrt{1 -D\sin^2\frac{\varphi+\psi_s(k_y) + \psi_{\bar{s}}(-k_y)}{2}},
\end{align}
with $D = 4|\tilde{t}_0|^2/(1+|\tilde{t}_0|^2)^2$ the transparency. In terms of the angle $\theta$, this becomes
\begin{align}
F_s=-\frac{\Delta k_FW}{2\pi}\int^{\pi/2}_{-\pi/2}d\theta\;\cos\theta\;\sqrt{1 -D\sin^2\frac{\varphi + \psi_s(\theta)}{2}},
\end{align}
with $\bar{s} = -s$, $\bar{\theta} = \theta+\pi$, and
\begin{align}
\psi_s(\theta) = -2s\beta k_\fermi d\left[\cos\gamma\sin\theta + \sin\gamma\frac{\cos2\theta}{2\cos\theta}\right].
\end{align}\\

\section{Andreev bound-state energies}
Consider a Josephson junction where the superconductors are separated by an altermagnetic layer. The interfaces are positioned at $x=\pm d/2$, and the superconducting phase is $\varphi/2$ in the left superconductor and $-\varphi/2$ in the right superconductor. The Bogoliubov-de Gennes equations, which can be thought of as the Schrödinger equation for the Andreev bound-states in the junction, take the form
\begin{equation}
    \begin{pmatrix}
        H(x,y) & \Delta(x,y) \\ \Delta^\dagger(x,y) & -H^T(x,y)
    \end{pmatrix}\begin{pmatrix}
        u_\uparrow \\ u_\downarrow \\ v_\uparrow \\ v_\downarrow 
    \end{pmatrix} = E \begin{pmatrix}
        u_\uparrow \\ u_\downarrow \\ v_\uparrow \\ v_\downarrow 
    \end{pmatrix}, 
\end{equation}
where $E$ is the energy of an Andreev bound-state. We note in passing that the bottom right part of the matrix is sometimes written $-H^*(x,y)$ in the literature, where the complex conjugate acts on the Pauli matrix $\sigma_y$. However, since the wave vector can generally be complex, this notation could cause confusion when ultimately expressing the above matrix in terms of wavevectors and we therefore find it more clear to use the matrix transpose. The superconducting gap matrix is
\begin{equation}
\Delta(x,y)=\Delta i\sigma_y\Theta(|x|-d/2)=\begin{pmatrix}
    0&\Delta\\-\Delta&0
\end{pmatrix}\Theta(|x|-d/2).
\end{equation}
$H(x,y)=H_0(x,y)+H_{AM}(x,y)+U_0\delta(|x|-d/2)$ consists of the normal part, the altermagnetic part, and the interfaces, respectively, and they are given by
\begin{equation}
    H_0(x,y)=-\frac{\hbar^2}{2m}\nabla^2-\mu
\end{equation}
with $\mu=\hbar^2k_F^2/2m$ and
\begin{widetext}
\begin{equation}
    H_{AM}(x,y)=-\frac{\hbar^2}{2m}t_1\left[\partial_y \Theta(d/2-|x|)\partial_x + \partial_x \Theta(d/2-|x|)\partial_y \right]\sigma_z -\frac{\hbar^2}{m}t_2\left[\partial_y \Theta(d/2-|x|)\partial_y - \partial_x \Theta(d/2-|x|)\partial_x \right]\sigma_z.
\end{equation}
\end{widetext}
The altermagnet is characterized by the dimensionless parameters $t_1$ and $t_2$, which depend on the altermagnet strength $t_0$ and the angle $\gamma$ of the altermagnet/superconductor interface relative to the crystalline axes through
\begin{equation}
    t_1 = 2t_0\cos(2\gamma), \qquad t_2=t_0\sin(2\gamma).
\end{equation}
For $\gamma=0$, the altermagnet has pure $d_{xy}$ symmetry, while for $\gamma=\pi/4$ the symmetry is $d_{x^2-y^2}$. Note that the step functions have been inserted in between the partial derivative operators. This is done to make the Hamiltonian Hermitian \cite{morrow_prb_84}. Also note that the $\partial_y$ term in the first bracket can be pulled outside everything and the $\partial_y$ term in the second bracket does not have to be symmetrized, as it commutes with the step-function. There exist other choices for where to place the step function (or, more generally, any spatially varying factor, in semiconductor computations this is typically the mass $m$), but the one chosen here is the most common. The form used here is similar to \eg that in Ref. \cite{papaj_prb_23} Eq. (S3).\\

Inside the superconductor, the wave vectors and the energy are related by
\begin{equation}
    E^2-|\Delta|^2 = \left( \frac{\hbar^2}{2m}(k_x^2+k_y^2) - \mu \right)^2.
\end{equation}
The group velocity is 
\begin{equation}
    v_g = \frac{\hbar^2}{m}\bigg(\frac{\hbar^2}{2m}(k_x^2+k_y^2-k_F^2)\bigg)\frac{k_x}{E}.
\end{equation}
Electron excitations live above the Fermi surface such that $k_x^2+k_y^2-k_F^2>0$, while hole excitations live below the Fermi surface such that $k_x^2+k_y^2-k_F^2<0$. Assuming $E>0$, the propagation direction for electrons is along $k_x$ while the propagation direction for holes is along $-k_x$.
The allowed $x$-components for the wave vectors are
\begin{equation}
    \pm q_\nu = \pm \sqrt{\nu\frac{2m}{\hbar^2}\sqrt{E^2-|\Delta|^2}+k_F^2-k_y^2}
\end{equation}
where $\pm$ refers to the propagation direction for electrons and $\nu=\text{sgn}(\varepsilon_q) = \text{sgn}((q^\nu)^2+k_y^2-k_F^2)$ is $+1$ for electrons and $-1$ for holes. For a more clean notation, we write $q_+=q_e$ and $q_-=q_h$. For holes, the propagation direction has the opposite sign of $q_h$ as mentioned previously.
In the right superconductor, the wave function is 
\begin{widetext}
\begin{equation}
\begin{split}
    \psi_{SC,R}(x,y)=e^{ik_yy} \bigg[ A_R\begin{pmatrix}
        u\\0\\0\\ve^{i\varphi/2}
    \end{pmatrix}e^{iq_e(x-d/2)} + B_R \begin{pmatrix}
        v\\0\\0\\ue^{i\varphi/2}
    \end{pmatrix}e^{-iq_h(x-d/2)} \\+ C_R \begin{pmatrix}
        0\\u\\-ve^{i\varphi/2}\\0
    \end{pmatrix}e^{iq_e(x-d/2)} + D_R \begin{pmatrix}
        0\\v\\-ue^{i\varphi/2}\\0
    \end{pmatrix}e^{-iq_h(x-d/2)} \bigg]
\end{split}
\end{equation}
where $u=\sqrt{\frac{1}{2}+\frac{1}{2}\frac{\sqrt{E^2-|\Delta|^2}}{E}}$ and $v=\sqrt{\frac{1}{2}-\frac{1}{2}\frac{\sqrt{E^2-|\Delta|^2}}{E}}$. This corresponds to eqs. C2 and C4 in Ref. \cite{sun_prb_23} except that the order parameter there is real ($\varphi=0$) (the similarity is seen by setting $g=1$ in Ref. \cite{sun_prb_23}). 
We note that the $e^{\pm iq_{e/h}L/2}$-factors in $\psi_{SC,R}(x,y)$ could be absorbed into the prefactors $A_R, B_R, C_R, D_R$. However, we keep $x-L/2$ explicitly because then the exponential factors $e^{\pm iq_{e/h} (x-L/2)}=1$ at the interface. This gives fewer $E$-dependent terms in the matrix determining the boundary conditions, which is an advantage. 
The $e^{ik_yy}$ factor is the same for electrons and holes because we have translational invariance in the $y$-direction, meaning that $k_y$ must be conserved.  In the left superconductor, the wave function is 
\begin{equation}
\begin{split}
    \psi_{SC,L}(x,y)=e^{ik_yy} \Big[ A_L\begin{pmatrix}
        u\\0\\0\\ve^{-i\varphi/2}
    \end{pmatrix}e^{-iq_e(x+d/2)} + B_L \begin{pmatrix}
        v\\0\\0\\ue^{-i\varphi/2}
    \end{pmatrix}e^{iq_h(x+d/2)} \\+ C_L \begin{pmatrix}
        0\\u\\-ve^{-i\varphi/2}\\0
    \end{pmatrix}e^{-iq_e(x+d/2)} + D_L \begin{pmatrix}
        0\\v\\-ue^{-i\varphi/2}\\0
    \end{pmatrix}e^{iq_h(x+d/2)} \Big].
\end{split}
\end{equation}
The difference between the left and right superconductor is only the prefactors, the sign of $\varphi$ and the propagation direction of the quasiparticles. \\

In the altermagnet, the Hamiltonian is diagonal. The allowed wave vectors are
\begin{equation}\label{eq:wavevectors}
    k_{\nu\sigma}^{\pm} = \frac{\pm \sqrt{(1-2\sigma t_2)\frac{2m}{\hbar^2}(\mu+\nu E)-(1-4t_2^2-t_1^2)k_y^2}-\sigma t_1 k_y}{1-2\sigma t_2},
\end{equation}
where $\nu$ refers to electrons/holes as before and $\sigma$ is spin. $\pm$ is the propagation direction for an electron as long as $t_1$ and $t_2$ are small. This is seen by calculating the group velocity in the $x$-direction: 
\begin{equation}
    v_g^{e\sigma} = \frac{\hbar^2}{m}(k_x+\sigma t_1 k_y-2\sigma t_2 k_x).
\end{equation}
If $t_1, t_2$ are large, the propagation direction of the electron is not necessarily in the same direction as $k_x$ anymore.
$ k_{\nu\sigma}^{\pm}$ is the same as eq. S10 in Ref. (\cite{papaj_prb_23}) (seen by using $t_{J1}/t_0=2t_2$, $t_{J2}/t_0=t_1$, $t_0=\hbar^2/2m$). The wavefunction in the altermagnet is 
\begin{equation}
\begin{split}
    \psi_{AM}(x,y)=e^{ik_yy}\bigg[&\left( a_\uparrow e^{ik_{e\uparrow}^+x} + b_\uparrow e^{ik_{e\uparrow}^-x} \right)\begin{pmatrix}
        1\\0\\0\\0
    \end{pmatrix}
     +\left( a_\downarrow e^{ik_{e\downarrow}^+x} + b_\downarrow e^{ik_{e\downarrow}^-x} \right)\begin{pmatrix}
        0\\1\\0\\0
    \end{pmatrix}
    \\+ &\left( c_\downarrow e^{ik_{h\uparrow}^+x} + d_\downarrow e^{ik_{h\uparrow}^-x} \right)\begin{pmatrix}
        0\\0\\1\\0
    \end{pmatrix}
    + \left( c_\uparrow e^{ik_{h\downarrow}^+x} + d_\uparrow e^{ik_{h\downarrow}^-x} \right)\begin{pmatrix}
        0\\0\\0\\1
    \end{pmatrix}\bigg].
\end{split}
\end{equation}
At each interface, probability and currents should be conserved. The boundary conditions demanding continuity of probability are
\begin{align}
    \psi_{SC,L}(-d/2) &= \psi_{AM}(-d/2)\\
    \psi_{SC,R}(+d/2) &= \psi_{AM}(+d/2).
\end{align}
The boundary conditions demanding
current conservation are obtained by integrating the Bogoliubov-de-Gennes equations on a small interval across the boundary at $x=L/2$, and they are given by
\begin{equation}
    \tau_z\otimes I(1-2t_2 I\otimes \sigma_z)\partial_x \psi_{AM}(d/2) + \tau_z\otimes[t_1\sigma_z \partial_y  + I\frac{2mU_0}{\hbar^2}]\psi_{AM}(d/2)=\tau_z\otimes I \partial_x \psi_{SC, R}(d/2)
\end{equation}
where $I$ is the $2\times 2$ identity matrix, $\tau$ are the Pauli matrices in Nambu (electron/hole) space and $\sigma$ are the Pauli matrices in spin space. At the left interface, the boundary condition is obtained by setting $\partial_x \rightarrow -\partial_x$, $d\rightarrow-d$, $t_1\rightarrow -t_1$ [because the $t_1$ term comes from $(\partial_x\Theta(x))\psi(x)$ which changes sign when $\Theta(x)\rightarrow \Theta(-x)$] and multiplying both sides with $-1$:
\begin{equation}
    \tau_z\otimes I(1-2t_2 I\otimes \sigma_z)\partial_x \psi_{AM}(-d/2) - \tau_z\otimes[-t_1\sigma_z \partial_y  + I\frac{2mU_0}{\hbar^2}]\psi_{AM}(-d/2)=\tau_z\otimes I \partial_x \psi_{SC, L}(-d/2).
\end{equation}
\\
To find the Andreev bound state energies, it is not necessary to compute all the prefactors in the wave functions. Therefore, we can use the boundary conditions to set up a system of equations on the form $A\boldsymbol{v}=\boldsymbol{0}$, where $A$ is a $16\times 16$ matrix and $\boldsymbol{v}$ contains all the prefactors $A_R, B_R, a_\uparrow$, and so on. We then proceed to solve the equation $\text{det}(A)=0$ for the energy. From the form of the wave functions, we see that the system of equations can be split up in two blocks: one for the first and fourth element in the wave functions, and one for the two middle elements. This is due to the absence of spin-mixing, and allows us to reduce the problem to computing the determinant of two $8\times8$ matrices. Consider first the first and fourth elements:
\begin{equation}
\begin{split}
    \psi_{AM}^\uparrow(x,y)=e^{ik_yy}\bigg[&\left( a_\uparrow e^{ik_{e\uparrow}^+x} + b_\uparrow e^{ik_{e\uparrow}^-x} \right)\begin{pmatrix}
        1\\0
    \end{pmatrix}
    + \left( c_\uparrow e^{ik_{h\downarrow}^+x} + d_\uparrow e^{ik_{h\downarrow}^-x} \right)\begin{pmatrix}
        0\\1
    \end{pmatrix}\bigg],
\end{split}
\end{equation}
\begin{equation}
    \psi_{SC,L}^\uparrow(x,y)=e^{ik_yy} \left[ A_L\begin{pmatrix}
        u\\ve^{-i\varphi/2}
    \end{pmatrix}e^{-iq_e(x+d/2)} + B_L \begin{pmatrix}
        v\\ue^{-i\varphi/2}
    \end{pmatrix}e^{iq_h(x+d/2)}\right]
\end{equation}
\begin{equation}
    \psi_{SC,R}^\uparrow(x,y)=e^{ik_yy} \left[ A_R\begin{pmatrix}
        u\\ve^{i\varphi/2}
    \end{pmatrix}e^{iq_e(x-d/2)} + B_R \begin{pmatrix}
        v\\ue^{i\varphi/2}
    \end{pmatrix}e^{-iq_h(x-d/2)} \right]
\end{equation}
Continuity of the wave function at the left interface gives
\begin{equation}
\begin{split}
    \begin{pmatrix}
        0\\0    \end{pmatrix}&=\psi_{SC,L}^\uparrow(-d/2)-\psi_{AM}^\uparrow(-d/2) %
        \\&=e^{ik_yy} \left[ \begin{pmatrix}
        A_Lu\\A_Lve^{-i\varphi/2}
    \end{pmatrix} + \begin{pmatrix}
        B_Lv\\B_Lue^{-i\varphi/2}
    \end{pmatrix} - \begin{pmatrix}
         a_\uparrow e^{ik_{e\uparrow}^+(-d/2)} + b_\uparrow e^{ik_{e\uparrow}^-(-d/2)} \\  c_\uparrow e^{ik_{h\downarrow}^+(-d/2)} + d_\uparrow e^{ik_{h\downarrow}^-(-d/2)}
    \end{pmatrix}
   \right] %
   \\&=e^{ik_yy} \left[ \begin{pmatrix}
        A_Lu \\A_Lve^{-i\varphi/2}
    \end{pmatrix} + \begin{pmatrix}
        B_Lv\\B_Lue^{-i\varphi/2}
    \end{pmatrix} - \begin{pmatrix}
         a_\uparrow e^{-ik_{e\uparrow}^+d/2} + b_\uparrow e^{-ik_{e\uparrow}^-d/2} \\  c_\uparrow e^{-ik_{h\downarrow}^+d/2} + d_\uparrow e^{-ik_{h\downarrow}^-d/2}
    \end{pmatrix}
    \right]
\end{split}
\end{equation}
Let the vector with the prefactors be defined as
\begin{equation}
    \boldsymbol{v}_\uparrow = \begin{pmatrix}
        A_L&B_L&a_\uparrow & b_\uparrow & c_\uparrow & d_\uparrow & A_R & B_R
    \end{pmatrix}^T,
\end{equation}
such that $A_\uparrow\boldsymbol{v}_\uparrow=\boldsymbol{0}$. The first row of $A_\uparrow$ is
\begin{equation}
    A_\uparrow^1=\begin{pmatrix}
        u  & v &  -e^{-ik_{e\uparrow}^+d/2} & -e^{-ik_{e\uparrow}^-d/2} &0&0&0&0
    \end{pmatrix}
\end{equation}
and the second row is
\begin{equation}
    A_\uparrow^2=\begin{pmatrix}
        ve^{-i\varphi/2} & ue^{-i\varphi/2} & 0 & 0 & -e^{-ik_{h\downarrow}^+d/2} & -e^{-ik_{h\downarrow}^-d/2} &0&0
    \end{pmatrix}.
\end{equation}
Continuity at the right interface gives
\begin{equation}
\begin{split}
    \begin{pmatrix}
        0\\0    \end{pmatrix}&=\psi_{SC,R}^\uparrow(d/2)-\psi_{AM}^\uparrow(d/2) %
        \\ &= e^{ik_yy} \left[ \begin{pmatrix}
        A_Ru\\A_Rve^{i\varphi/2}
    \end{pmatrix} + \begin{pmatrix}
        B_R v \\B_R ue^{i\varphi/2} 
    \end{pmatrix}
    - \begin{pmatrix}
         a_\uparrow e^{ik_{e\uparrow}^+d/2} + b_\uparrow e^{ik_{e\uparrow}^-d/2} \\  c_\uparrow e^{ik_{h\downarrow}^+d/2} + d_\uparrow e^{ik_{h\downarrow}^-d/2}
    \end{pmatrix}
    \right].
\end{split}
\end{equation}
This gives
\begin{equation}
    A_\uparrow^3=\begin{pmatrix}
       0&0 &  -e^{ik_{e\uparrow}^+d/2} & -e^{ik_{e\uparrow}^-d/2} &0&0& u  & v
    \end{pmatrix}
\end{equation}
\begin{equation}
    A_\uparrow^4=\begin{pmatrix}
        0&0 & 0 & 0 & -e^{ik_{h\downarrow}^+L/2} & -e^{ik_{h\downarrow}^-d/2} &ve^{i\varphi/2} & ue^{i\varphi/2}
    \end{pmatrix}.
\end{equation}
The next is current conservation. We have
\begin{equation}
\begin{split}
   [ \begin{pmatrix}
        1&0\\0&-1
    \end{pmatrix}-2t_2 \begin{pmatrix}
        1&0\\0&1
    \end{pmatrix}]\partial_x \psi_{AM}^\uparrow(d/2) + [ \begin{pmatrix}
        1&0\\0&1
    \end{pmatrix}t_1 \partial_y  +  \begin{pmatrix}
        1&0\\0&-1
    \end{pmatrix}\frac{2mU_0}{\hbar^2}]\psi_{AM}^\uparrow(d/2)- \begin{pmatrix}
        1&0\\0&-1
    \end{pmatrix} \partial_x \psi_{SC, R}^\uparrow(d/2)=\begin{pmatrix}
        0\\0
    \end{pmatrix}
\end{split}
\end{equation}
\begin{equation*}
\begin{split}
    \begin{pmatrix}
        1-2t_2&0\\0&-1-2t_2
    \end{pmatrix}
    e^{ik_yy}\begin{pmatrix}
        a_\uparrow ik_{e\uparrow}^+ e^{ik_{e\uparrow}^+d/2} + b_\uparrow ik_{e\uparrow}^-e^{ik_{e\uparrow}^-d/2}\\ c_\uparrow ik_{h\downarrow}^+ e^{ik_{h\downarrow}^+d/2} + d_\uparrow ik_{h\downarrow}^- e^{ik_{h\downarrow}^-d/2}\end{pmatrix} + 
        \begin{pmatrix}
        t_1 ik_y+\frac{2mU_0}{\hbar^2}&0\\0&t_1 ik_y-\frac{2mU_0}{\hbar^2}
    \end{pmatrix}e^{ik_yy}\begin{pmatrix}
        a_\uparrow e^{ik_{e\uparrow}^+d/2} + b_\uparrow e^{ik_{e\uparrow}^-d/2}\\ c_\uparrow e^{ik_{h\downarrow}^+d/2} + d_\uparrow e^{ik_{h\downarrow}^-d/2}\end{pmatrix}  \\- \begin{pmatrix}
        1&0\\0&-1
    \end{pmatrix}e^{ik_yy} \left[ \begin{pmatrix}
        A_Ruiq_e\\A_Rviq_e e^{i\varphi/2}
    \end{pmatrix} + \begin{pmatrix}
        B_Rv(-iq_h)\\B_Ru(-iq_h)e^{i\varphi/2}
    \end{pmatrix} \right] 
    =\begin{pmatrix}
        0\\0
    \end{pmatrix}
\end{split}
\end{equation*}
\begin{equation*}
\begin{split}
    \begin{pmatrix}
        1-2t_2&0\\0&-1-2t_2
    \end{pmatrix}
    \begin{pmatrix}
        a_\uparrow k_{e\uparrow}^+ e^{ik_{e\uparrow}^+d/2} + b_\uparrow k_{e\uparrow}^-e^{ik_{e\uparrow}^-d/2}\\ c_\uparrow k_{h\downarrow}^+ e^{ik_{h\downarrow}^+d/2} + d_\uparrow k_{h\downarrow}^- e^{ik_{h\downarrow}^-d/2}\end{pmatrix} + 
        \begin{pmatrix}
        t_1 k_y-i\frac{2mU_0}{\hbar^2}&0\\0&t_1 k_y+i\frac{2mU_0}{\hbar^2}
    \end{pmatrix}\begin{pmatrix}
        a_\uparrow e^{ik_{e\uparrow}^+d/2} + b_\uparrow e^{ik_{e\uparrow}^-d/2}\\ c_\uparrow e^{ik_{h\downarrow}^+d/2} + d_\uparrow e^{ik_{h\downarrow}^-d/2}\end{pmatrix}  \\+ \begin{pmatrix}
        -1&0\\0&1
    \end{pmatrix} \left[ \begin{pmatrix}
        A_Ruq_e\\A_Rvq_e e^{i\varphi/2}
    \end{pmatrix} + \begin{pmatrix}
        -B_Rvq_h\\-B_Ruq_he^{i\varphi/2}
    \end{pmatrix} \right] 
    =\begin{pmatrix}
        0\\0
    \end{pmatrix}.
\end{split}
\end{equation*}
Multiplying the equation with the coherence length $\xi$ makes it dimensionless. This gives 
\begin{equation}
    A^5_\uparrow=\begin{pmatrix}
        0 \\ 0 \\ (1-2t_2)\xi k_{e\uparrow}^+e^{ik_{e\uparrow}^+d/2} + (t_1 \xi k_y-i\frac{2m \xi U_0}{\hbar^2})e^{ik_{e\uparrow}^+d/2}
        \\(1-2t_2)\xi k_{e\uparrow}^-e^{ik_{e\uparrow}^-d/2} + (t_1 \xi k_y-i\frac{2m\xi U_0}{\hbar^2})e^{ik_{e\uparrow}^-d/2}
        \\0 \\0 \\ -u\xi q_e \\ v\xi q_h
    \end{pmatrix}^T
\end{equation}
\begin{equation}
    A^6_\uparrow = \begin{pmatrix}
        0\\0\\0\\0\\
        (-1-2t_2) \xi k_{h\downarrow}^+ e^{ik_{h\downarrow}^+d/2} +(t_1\xi k_y+i\frac{2m\xi U_0}{\hbar^2})e^{ik_{h\downarrow}^+d/2}
        \\
        (-1-2t_2)\xi k_{h\downarrow}^- e^{ik_{h\downarrow}^-
        d/2}+(t_1 \xi k_y+i\frac{2mU_0}{\hbar^2})e^{ik_{h\downarrow}^-d/2}
        \\
        v\xi q_e e^{i\varphi/2}
        \\
        -u\xi q_he^{i\varphi/2}
    \end{pmatrix}^T.
\end{equation}

The current conserving boundary condition on the left interface is
\begin{equation}
\begin{split}
   [ \begin{pmatrix}
        1&0\\0&-1
    \end{pmatrix}-2t_2 \begin{pmatrix}
        1&0\\0&1
    \end{pmatrix}]\partial_x \psi_{AM}^\uparrow(-d/2) - [ \begin{pmatrix}
        1&0\\0&1
    \end{pmatrix}(-t_1) \partial_y  +  \begin{pmatrix}
        1&0\\0&-1
    \end{pmatrix}\frac{2mU_0}{\hbar^2}]\psi_{AM}^\uparrow(-d/2)- \begin{pmatrix}
        1&0\\0&-1
    \end{pmatrix} \partial_x \psi_{SC, L}^\uparrow(-d/2)=\begin{pmatrix}
        0\\0
    \end{pmatrix}
\end{split}
\end{equation}
\begin{equation*}
\begin{split}
    \begin{pmatrix}
        1-2t_2&0\\0&-1-2t_2
    \end{pmatrix}
    \begin{pmatrix}
        a_\uparrow ik_{e\uparrow}^+ e^{-ik_{e\uparrow}^+d/2} + b_\uparrow ik_{e\uparrow}^-e^{-ik_{e\uparrow}^-d/2}\\ c_\uparrow ik_{h\downarrow}^+ e^{-ik_{h\downarrow}^+d/2} + d_\uparrow ik_{h\downarrow}^- e^{-ik_{h\downarrow}^-d/2}\end{pmatrix} + 
        \begin{pmatrix}
        t_1 ik_y-\frac{2mU_0}{\hbar^2}&0\\0&t_1 ik_y+\frac{2mU_0}{\hbar^2}
    \end{pmatrix}\begin{pmatrix}
        a_\uparrow e^{-ik_{e\uparrow}^+d/2} + b_\uparrow e^{-ik_{e\uparrow}^-d/2}\\ c_\uparrow e^{-ik_{h\downarrow}^+d/2} + d_\uparrow e^{-ik_{h\downarrow}^-d/2}\end{pmatrix}  \\- \begin{pmatrix}
        1&0\\0&-1
    \end{pmatrix} \left[ \begin{pmatrix}
        A_Lu(-iq_e)\\A_Lv(-iq_e)e^{-i\varphi/2}
    \end{pmatrix} +  \begin{pmatrix}
        B_Lv(iq_h)\\B_Lu(iq_h)e^{-i\varphi/2}
    \end{pmatrix}\right] 
    =\begin{pmatrix}
        0\\0
    \end{pmatrix}
\end{split}
\end{equation*}
\begin{equation*}
\begin{split}
    \begin{pmatrix}
        1-2t_2&0\\0&-1-2t_2
    \end{pmatrix}
    \begin{pmatrix}
        a_\uparrow k_{e\uparrow}^+ e^{-ik_{e\uparrow}^+d/2} + b_\uparrow k_{e\uparrow}^-e^{-ik_{e\uparrow}^-d/2}\\ c_\uparrow k_{h\downarrow}^+ e^{-ik_{h\downarrow}^+d/2} + d_\uparrow k_{h\downarrow}^- e^{-ik_{h\downarrow}^-d/2}\end{pmatrix} + 
        \begin{pmatrix}
        t_1 k_y+i\frac{2mU_0}{\hbar^2}&0\\0&t_1 k_y-i\frac{2mU_0}{\hbar^2}
    \end{pmatrix}\begin{pmatrix}
        a_\uparrow e^{-ik_{e\uparrow}^+d/2} + b_\uparrow e^{-ik_{e\uparrow}^-d/2}\\ c_\uparrow e^{-ik_{h\downarrow}^+d/2} + d_\uparrow e^{-ik_{h\downarrow}^-d/2}\end{pmatrix}  \\+ \begin{pmatrix}
        -1&0\\0&1
    \end{pmatrix} \left[ \begin{pmatrix}
        -A_Luq_e\\-A_Lvq_ee^{-i\varphi/2}
    \end{pmatrix} +  \begin{pmatrix}
        B_Lvq_h\\B_Luq_he^{-i\varphi/2}
    \end{pmatrix}\right] 
    =\begin{pmatrix}
        0\\0
    \end{pmatrix}.
\end{split}
\end{equation*}
This gives
\begin{equation}
    A_\uparrow^7=\begin{pmatrix}
        u\xi q_e\\
        -v\xi q_h\\
        (1-2t_2)\xi k_{e\uparrow}^+ e^{-ik_{e\uparrow}^+d/2} +(t_1\xi k_y+i\frac{2m\xi U_0}{\hbar^2})e^{-ik_{e\uparrow}^+d/2}
        \\
        (1-2t_2) \xi k_{e\uparrow}^- e^{-ik_{e\uparrow}^-d/2}+(t_1 \xi k_y+i\frac{2m\xi U_0}{\hbar^2})e^{-ik_{e\uparrow}^-d/2}
        \\0\\0\\0\\0
    \end{pmatrix}^T
\end{equation}
\begin{equation}
    A_\uparrow^8=\begin{pmatrix}
        -v\xi q_ee^{-i\varphi/2}\\
        u \xi q_he^{-i\varphi/2}\\
        0\\
        0\\
        (-1-2t_2)\xi k_{h\downarrow}^+ e^{-ik_{h\downarrow}^+d/2} +(t_1 \xi k_y-i\frac{2m\xi U_0}{\hbar^2})e^{-ik_{h\downarrow}^+d/2} \\
        (-1-2t_2)\xi k_{h\downarrow}^- e^{-ik_{h\downarrow}^-d/2} +(t_1\xi k_y-i\frac{2m\xi U_0}{\hbar^2})e^{-ik_{h\downarrow}^-d/2}\\
        0\\
        0
    \end{pmatrix}^T.
\end{equation}

There is a similar set of equations for the second and third elements in the wave functions. These equations are found by setting $e^{i\varphi/2}\rightarrow -e^{i\varphi/2}$ and $k_{e/h,\sigma}^\pm\rightarrow k_{e/h,-\sigma}^\pm$ in $A_\downarrow$. Moreover, due to the boundary conditions, we must also set all explicit entries of $t_1 \to -t_1$ and $t_2 \to -t_2$ (but not change these signs in the wavevectors in the altermagnetic region). As a consistency check, we have verified that we are able to reproduce the results of \cite{beenakker_prb_23} in various limiting cases.

The interface transparency is characterized by $Z=mU_0/(\hbar^2k_F)$. $Z=0$ is a perfect interface while a tunneling interface can be modeled by $Z\sim3$.

\subsection{Calculation of the free energy and the current-phase relation}
Inside the altermagnet, $k_x$ and $k_y$ should be real to have propagating waves. Therefore, the square root in $k_{\nu\sigma}^\pm$ should be real, and the expression inside the square root should be positive. The expression depends on $E$, but in the limit $\mu \gg E$ we can ignore this energy dependence, such that
\begin{equation}
    k_y^{\text{max},\sigma} = k_F \sqrt{\frac{1-2\sigma t_2}{1-4t_2^2-t_1^2}}.
\end{equation}
Since we can have Andreev reflections where $e\uparrow$ can be reflected as $h\downarrow$ and $k_y$ is conserved in this process, all $k_y$ values must be valid for both spins. Since $k_y^{\text{max}, \downarrow} > k_y^{\text{max},\uparrow}$, $\sigma=\uparrow$ sets the upper limit on $k_y$. Therefore, $k_y\in[-k_y^{\text{max},\uparrow}, k_y^{\text{max},\uparrow}]$.

The Josephson junction has a finite width $W$, and we assume that $W$ is so large that any edge-effects on the supercurrent are negligible. In this case, the sum over the $k_y$-values in the free energy can be turned into an integral times the width of the junction. \\

At zero temperature, the free energy is given by
\begin{equation}\label{eq:AM free energy}
\begin{split}
    F &= \sum_{ky, E<0} E = \frac{W}{2\pi} \sum_{n=-N}^{N} E \frac{2\pi}{W} = \frac{W}{2\pi} \int_{-k_y^{max}}^{k_y^{max}}dk_y E = \frac{W/\xi|\Delta|}{2\pi} \sum_{\xi k_y}\frac{E}{|\Delta|} \frac{2\xi k_y^{max}}{N_y} \\&= \frac{|\Delta|W/\xi \cdot \xi k_y^{max}}{\pi} \cdot \frac{1}{N_y} \sum_{|\xi k_y| < \xi k_y^{max}}\sum_{-|\Delta|<E<0} \frac{E}{|\Delta|}.  
\end{split}
\end{equation}
\end{widetext}

\section{Origin of dielectric noise}\label{sec:appendixdielectric}
The treatment of charge and flux noise in the qubit is rather clear since they are associated with the variation of the charge density on the capacitor and the flux penetrating the ring in the flux qubit case, respectively. Hence, the noise parameter $\lambda$ which enters the analytical expressions for the computation of decoherence in the main text is $n_g$ and $\Phi$ in those cases. Which parameter to choose as $\lambda$ in the case of dielectric noise is less obvious at first glance, and rarely discussed in particular detail in recent literature. Hence, we include a short exposition of this here.

A qubit typically has several electrical circuit components where some of these components may be grown on substrates. This is the case for Josephson junctions, which typically are grown on top of a substrate, for instance in a planar geometry. This substrate may in a realistic experimental setting contain defects, which can be modelled in an approximative manner as two-level quantum systems (TLS) that emit and absorb photons through electromagnetic fluctuations. When this happens, it can cause current fluctuations through the Josephson element of the qubit, as charges are excited and de-excited. In particular, there can exist an electric dipole moment associated with the TLS which varies as the state of the TLS fluctuates by means of \eg electrons hopping between the two states. The presence of an electric dipole moment is going to affect the potential experienced by Cooper pairs tunneling through the Josephson junction \cite{simmonds_prl_04, martinis_prl_05}. This causes the TLS to modify the already existing tunneling barrier between the superconductors in the junction. Since the current depends sensitively on the tunneling probability, the fluctuating TLS causes variations in the current. In other words, dielectric noise leads to current noise. \\

This can be modelled in a qubit Hamiltonian as follows. First, we have to account for the fact that in the presence of an applied current bias $I$ (as a fixed condition) to the system, there is an extra term in the Hamiltonian. The total current in a Josephson junction in general consists of a supercurrent and resistive contribution:
\begin{align}
I = I_c\sin\varphi + I_N,\; I_N= \frac{V}{R}
\end{align}
where $V$ is the voltage difference between the superconductors. It is related to the gauge-invariant superconducting phase difference $\varphi$ as
\begin{align}
V = \frac{\hbar\dot{\varphi}}{2e}
\end{align}
The Joule heating $I_NV$ corresponds to the power dissipated by the resistor in the system.  

Now, the total time derivative of the system energy will be the power dissipated in the system. For a free energy $F$ that depends on the superconducting phase difference, we have
\begin{align}
\frac{d F}{dt} = \frac{\partial F}{\partial t} +  \frac{\partial F}{\partial \varphi}\frac{\partial\varphi}{\partial t} 
\end{align}
Assume that there is no explicit time-dependence in the Hamiltonian of our system. Then, the remaining term has to be the Joule heating power of the system.  Therefore, we have
\begin{align}
-I_NV = \frac{\hbar\dot{\varphi}}{2e}[I - I_c\sin(\phi)] = \frac{\partial F}{\partial \varphi}\frac{\partial\varphi}{\partial t}.
\end{align}
This can be rewritten as
\begin{align}
-\frac{\hbar}{2e}[I - I_c\sin(\phi)] = \frac{\partial F}{\partial \varphi}
\end{align}
which we can solve for the free energy $F$. We obtain
\begin{align}\label{eq:freenophi0}
F = \text{constant} - \frac{\hbar}{2e}(I\varphi + I_c\cos\varphi)
\end{align}
We see the regular Josephson energy and the additional $I\varphi$ term present. Thus, the Hamiltonian has acquired an extra term
$-\frac{\hbar I}{2e}\varphi$.
We can set the current bias to zero, and consider the effect of fluctuations $\delta I$. The Hamiltonian can be expanded in terms of a noise parameter $\lambda$: $H=H_0+\partial_\lambda H \cdot \delta\lambda(t)=H_0+V_\lambda(t)$, with $\lambda=\lambda_0+\delta\lambda(t)$ and $\partial_\lambda H$ is short for $\partial_\lambda H(\lambda)|_{\lambda=\lambda_0}$. Note that $H_0 = H(\lambda_0)$. The noise is in the current, so $\lambda = I = I_0 + \delta I$ with $I_0=0$. In this case, it is clear that $\partial_{I} H = - \frac{\hbar\varphi}{2e}$, which is precisely the operator entering the matrix element for dielectric noise.

\section{Dephasing from dielectric noise}\label{sec:appendixdephasing}

Finally, we show why dephasing due to dielectric noise in the form of current-fluctuations can be disregarded in our altermagnetic transmon. The general argument goes as follows. Consider a qubit Hamiltonian perturbed by a classical, stationary, zero-mean noise process $\delta\lambda(t)$ that couples linearly to some operator $\hat{X}$ of the qubit,
\begin{equation}
H(t) = H_0 + A\,\hat{X}\, \delta\lambda(t),
\label{eq:noise_hamiltonian}
\end{equation}
where $H_0$ is the unperturbed Hamiltonian with eigenstates $\{\ket{0},\ket{1},\dots\}$ and eigenenergies $\{E_0,E_1,\dots\}$, $A$ is a coupling constant, and $\hat{X}$ may represent \eg the charge operator $\hat n$, the phase operator $\hat\varphi$, or the flux operator $\hat\Phi$, depending on the physical noise channel under consideration (charge noise, dielectric noise, flux noise). Restricting to the computational subspace $\{\ket{0},\ket{1}\}$, the effective Hamiltonian becomes
\begin{equation}
H_{\mathrm{eff}}(t) =
\begin{pmatrix}
E_0 + A\,\delta\lambda(t)\,\bra{0}\hat{X}\ket{0} & A\,\delta\lambda(t)\,\bra{0}\hat{X}\ket{1} \\[4pt]
A\,\delta\lambda(t)\,\bra{1}\hat{X}\ket{0} & E_1 + A\,\delta\lambda(t)\,\bra{1}\hat{X}\ket{1}
\end{pmatrix}.
\label{eq:Heff_matrix}
\end{equation}
The off-diagonal and diagonal matrix elements of $\hat{X}$ play qualitatively different roles, leading respectively to relaxation and to dephasing.

The off-diagonal entries $\bra{0}\hat{X}\ket{1}$, $\bra{1}\hat{X}\ket{0}$ couple $\ket{0}$ and $\ket{1}$ through the time-dependent noise $\delta\lambda(t)$. Within the standard Fermi's-golden-rule treatment, the component of $\delta\lambda(t)$ oscillating at the qubit transition frequency $\omega = (E_1-E_0)/\hbar$ drives resonant transitions between $\ket{0}$ and $\ket{1}$, giving rise to energy relaxation with a rate
\begin{equation}
\Gamma_1 \propto |A|^2\,|\bra{0}\hat{X}\ket{1}|^2\,S_\lambda(\omega),
\label{eq:gamma1}
\end{equation}
where {$S_\lambda(\omega)$} is the power spectral density of the noise $\delta\lambda(t)$. A nonzero $\bra{0}\hat{X}\ket{1}$ is thus the necessary condition for this noise channel to contribute to $T_1$. This corresponds to Eq. (\ref{eq:relaxationrate}) for the relaxation rate in the main text.

The diagonal entries of Eq.~\eqref{eq:Heff_matrix} simply shift each bare energy level $E_k$ by an amount $A\,\delta\lambda(t)\,\bra{k}\hat{X}\ket{k}$. The instantaneous qubit splitting becomes
\begin{align}
\hbar\,\omega(t) &= \big[E_1 + A \delta\lambda(t)\bra{1}\hat{X}\ket{1}\big] - \big[E_0 + A \delta\lambda(t)\bra{0}\hat{X}\ket{0}\big]
\notag\\
&= \hbar\,\omega^{(0)} + A \delta\lambda(t)\Big(\bra{1}\hat{X}\ket{1}-\bra{0}\hat{X}\ket{0}\Big).
\label{eq:omega01_fluctuation}
\end{align}
Equivalently, the fluctuation of the transition frequency is
\begin{align}
\delta\omega(t) &= \big(\partial_\lambda \omega\big)\,\delta\lambda(t),
\notag\\
\hbar\,\partial_\lambda\omega &\equiv A\Big(\bra{1}\hat{X}\ket{1}-\bra{0}\hat{X}\ket{0}\Big).
\label{eq:domega_def}
\end{align}
The key observation is that only the difference of the diagonal matrix elements enters $\delta\omega(t)$. If the noise shifts both levels by the same amount, i.e.\ $\bra{1}\hat{X}\ket{1}=\bra{0}\hat{X}\ket{0}$, this is a common-mode energy shift that leaves the transition frequency $\omega$ unchanged. Since dephasing arises from fluctuations of the \emph{relative} phase accumulated between $\ket0$ and $\ket1$, such a common-mode shift has no effect on the qubit coherence to first order. A nonzero diagonal-element difference is therefore the necessary condition for first-order pure dephasing.
We can now see why the pure-dephasing contribution from this channel ($\hat{X} = \hat{\varphi}$) vanishes in our case. Whenever the eigenstates $\ket{0}$, $\ket{1}$ have definite parity under $\hat\varphi\to-\hat\varphi$, then since $\hat\varphi$ is an odd operator under this symmetry, $\bra{k}\hat{\varphi}\ket{k}=0$ for all $k$. This is the case in our system since our potential Eq. (\ref{eq:potentialcos}) only has cos-terms and since $I=0$. In turn, this means $\bra{1}\hat{\varphi}\ket{1}-\bra{0}\hat{\varphi}\ket{0}=0$ and $\partial_\lambda\omega_{01}=0$ identically. No first-order dephasing thus arises from this channel regardless of the noise spectrum.

A similar argument applies to charge noise at the zero-offset-charge
sweet spot. For
\[
H=4E_c(\hat n-n_g)^2+U(\hat\varphi),
\]
charge fluctuations couple through
\[
\left.\partial_{q_g}H\right|_{n_g=0}=-\frac{8E_c}{2e}\hat n .
\]
At \(n_g=0\), and for a potential satisfying \(U(\varphi)=U(-\varphi)\),
the Hamiltonian is invariant under \(\varphi\to-\varphi\). Its eigenstates
can therefore be chosen to have definite parity. Since
\(\hat n=-i\partial_\varphi\) is odd under this parity operation,
\[
\bra{k}\hat n\ket{k}=0
\]
for every eigenstate \(k\). Consequently,
\[
\left.\partial_{q_g}\omega_{01}\right|_{n_g=0}=0,
\]
and charge noise does not give rise to first-order pure dephasing at
zero offset charge. For \(n_g\neq0\), the term
\(-8E_C n_g\hat n\) breaks this parity symmetry, so this cancellation is
not generally present.

\bibliography{references}

\end{document}